\documentclass[12pt,a4paper,oneside]{article}

\usepackage{times}
\usepackage{amsmath, amssymb,rotating,fancybox,prodint}
\usepackage{dsfont}
\usepackage{graphicx}
\usepackage{leftindex}

\usepackage[latin1]{inputenc}
\usepackage[usenames]{color}
\usepackage[authoryear]{natbib}

\usepackage{enumitem}
\usepackage{longtable} \usepackage{booktabs}
\usepackage{nomencl}
\usepackage{multirow}

\newtheorem{theorem}{Theorem}
\newtheorem{lemma}{Lemma}

\newcounter{AnnahmenCounter}

\newcounter{AnnahmenCounterLocal}
\newcounter{AnnahmenCounterGlobal}
\newcounter{AnnahmenCounterA}
\newcounter{AnnahmenCounterB}

\newcounter{AnnahmenCounterP}
\newcounter{AnnahmenCounterZ}
\newcounter{AnnahmenCounterM}
\newcounter{AnnahmenCounterK}
\newcounter{AnnahmenCounterKa}

\newcommand{\AnnA}{A}
\newcommand{\AnnB}{B}

\newcommand{\AnnANumm}{(\AnnA\arabic{*})}

\newcommand{\AnnBNumm}{(\AnnB\arabic{*})}

\usepackage[]{graphicx}
\chardef\bslash=`\\ 

\def\qed{\space$\rule{0.1in}{0.1in}$ \par \vspace{.15in}}

\def\bSig\mathbf{\Sigma}

\title{Left-Truncated Health Insurance Claims Data: Theoretical Review and Empirical Application}

\author{R. Wei\ss{}bach$^1$, A. D\"{o}rre$^1$, D. Wied$^2$, G. Doblhammer$^{3,4}$, A. Fink$^4$ \\ {\footnotesize Chairs in Statistics and Econometrics, Universities of Rostock$^1$ and K\"oln$^2$} \\ {\footnotesize Chair in Empirical Social Research/Demography, University of Rostock$^3$ and DZNE, Bonn$^4$}}
\date{}

\begin{document}

\maketitle

\vspace*{-0.7cm}

\begin{abstract}
At the beginning of 2004, we draw a sample of size 0.25 million people from the inventory of the health insurer AOK. We followed their health claims until 2013. Our aim is the effect a stroke on the dementia onset probability, for Germans born in the first half of the 20$^{th}$ century. People deceased before 2004 are randomly left-truncated. Filtrations, modelling the missing data, enable to circumvent the unknown number of truncated persons by using a conditional instead of the full likelihood. Dementia onset after 2013 is a conditionally fixed right-censoring event. 
For each observed health history, Jacod's formula yields the conditional likelihood contribution. Asymptotic normality of the estimated intensities is derived, relative to a sample size definition that includes the truncated people. Yet, the standard error is observable. The claims data reveal that after a stroke, with time measured in years, the intensity of dementia onset increases from 0.02 to 0.07. Using the independence of the two estimated intensities, a 95\%-confidence interval for their difference is [0.050,0.056]. The effect halves, when we extend the analysis to an age-inhomogeneous model, but does not change further when we additionally adjust for multi-morbidity.
\end{abstract}

\section{Introduction} \label{introduc}

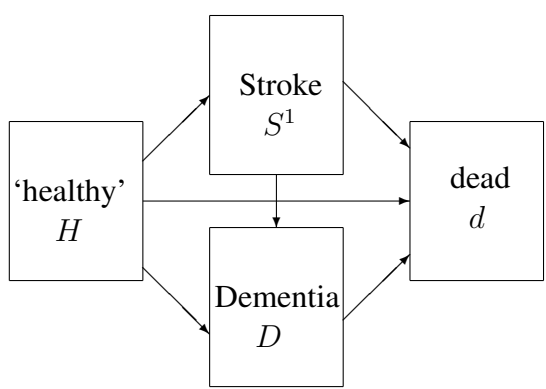
\begin{figure}[b]
	\begin{center}
		\begin{picture}(200, 130)
			\put(10.0, 95.0){\thinlines\line(0, -1){60.0}} 
			\put(10.0, 35.0){\thinlines\line(1, 0){50.0}} 
			\put(60.0, 35.0){\thinlines\line(0, 1){60.0}} 
			\put(60.0, 95.0){\thinlines\line(-1, 0){50.0}} 
			
			\put(85.0, -5.0){\thinlines\line(0, 1){60.0}} 
			\put(85.0, -5.0){\thinlines\line(1, 0){50.0}} 
			\put(135.0,	-5.0){\thinlines\line(0, 1){60.0}} 
			\put(135.0, 55.0){\thinlines\line(-1, 0){50.0}}
			
			\put(85.0, 75.0){\thinlines\line(0, 1){60.0}} 
			\put(85.0, 75.0){\thinlines\line(1, 0){50.0}} 
			\put(135.0,	75.0){\thinlines\line(0, 1){60.0}} 
			\put(135.0, 135.0){\thinlines\line(-1, 0){50.0}}  
			
			\put(160.0, 35.0){\thinlines\line(1, 0){50.0}} 
			\put(210.0,	35.0){\thinlines\line(0, 1){60.0}} 
			\put(160.0, 35.0){\thinlines\line(0, 1){60.0}} 
			\put(160.0, 95.0){\thinlines\line(1, 0){50.0}} 
			
			\put(110.0, 75.0){\thinlines\vector(0, -1){20.0}} 
			\put(135.0, 20.0){\thinlines\vector(1, 1){25.0}} 
			\put(135.0, 110.0){\thinlines\vector(1, -1){25.0}} 
			\put(60.0, 40.0){\thinlines\vector(1, -1){25.0}} 
			\put(60.0, 80.0){\thinlines\vector(1, 1){25.0}} 
			\put(60.0, 65.0){\thinlines\vector(1, 0){100.0}}

			\put(11.0, 65.0){`healthy'} 
				\put(27.0, 50.0){$H$} 
			\put(96.0, 105.0){Stroke}  
			\put(105.0, 90.0){$S^1$}  
			\put(87.0, 25.0){Dementia}  
			\put(102.0, 10.0){$D$}  
			\put(175.0, 70.0){dead} 
				\put(182.0, 55.0){$d$} 
		\end{picture}\caption{Disease states and transitions}\label{gesch}
	\end{center}
\end{figure}

For Germany, \cite{jessen2018} forecast an increase up to 2.8 million people with dementia in 2050.  One risk factor is a stroke and we model life as a time-continuous multi-state history (see Figure \ref{gesch}). The model is also called `disability model' \cite[see][Figure 1.6]{Hou}.
Note that `healthy' only stands synonymous for `neither having dementia nor having had a stroke'. Dementia onset of a person after a stroke (or precisely, after its first) is, ceteris paribus, governed by the onset intensity, named $\lambda_{S^1D}$. We want to compare the intensity to the dementia onset intensity of a healthy person, $\lambda_{HD}$. By stroke effect on dementia onset we refer to the difference (or the ratio) of these two intensities. 

\begin{figure}[b]
		\begin{picture}(300, 70)
			
			
			\put(50.0, 60.0){\Ovalbox{Population: Germans born 1900--1954 (Size $76{,}239{,}006$)}}
			\put(180.0, 53.0){\thinlines\vector(0, -1){10.0}}
			
			\put(65.0, 30.0){\ovalbox{Left-Truncated: People not surviving 31/12/2003}}
			\put(180.0, 23.0){\thinlines\vector(0, -1){10.0}}
			
			\put(47.0, 0.0){\ovalbox{Right-censored: Transitions to $S^1$,$D$ or $d$ past 31/12/2013}}
		\end{picture}
	\caption{Data: Trajectory from population to the left-truncated mass and right-censored persons (Population size: German Statistical Office (2004), without stillborn)}
	\label{data_selection}
\end{figure}
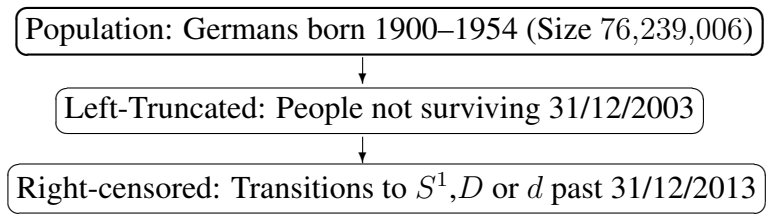

The population, we will infer to, are Germans born in the years 1900 to 1954, which we for simplicity occasionally call `the first half of the 20$^{th}$ century' (see Figure \ref{data_selection}, top box). Drawing a simple random sample (size {\em n}$_{all}$) from that population and then truncating, i.e. not observing, persons deceased before 2004, is similar to drawing a simple random sample (size {\em n}) of all Germans alive 2004. The latter is our situation. Note that in the first design, the number of observations is random, and in the second design the number {\em n}$_{all}$ is unknown. That people are missing in the data due to an early death is called left-truncation (see Figure \ref{data_selection}, middle box and Figures \ref{bspbild} or \ref{c_figure} in the Appendix) and is a typical design defect in disease state models \cite[see e.g.][]{putter2006}. Ignoring truncation would lead to the `immortal time bias' \cite[see e.g.][]{HERNAN201670,yadav2021}.

Efficient estimation in large samples is usually achieved by the maximum likelihood method. In order to render the knowledge about truncated persons obsolete, here marginalisation and conditioning are necessary. We intend both, a review of the methodological arguments and the application to a case study about the effect of stroke on dementia onset in Germany. Right-censoring will also be accounted for (see Figure \ref{data_selection}, bottom box), but has already a large literature. 
Explaining how the technique reduces information is easier when the state `being alive' is not subdivided into  several disease states, and is explained in a lifetime state model \cite[in the terminology of][]{Hou} in Appendix \ref{lsm}. Then, with all states  of Figure \ref{gesch}, Section \ref{dsm1} derives a confidence interval for the difference $\lambda_{S^1D}- \lambda_{HD}$ and a Wald-type test for the stroke effect. In the next Section \ref{sec3}, we allow intensities to dependent on age for two reasons. On the one hand, it allows to compare our case study with re-known international studies where age-inhomogeneous behaviour is routinely accounted for. On the other hand, we will see that accounting for age changes the stroke effect, drastically, and is an instructive example for Simpson's paradox. In Section \ref{sec3} we additionally adjust for (vascular) multi-morbidity, in order to answer the question whether the elevated dementia risk by stroke could be anticipated for multi-morbid persons. Appendix \ref{lsm}, Section \ref{dsm1} and Section \ref{dsm2} start with the asymptotic theory, Appendix \ref{lsm} and Section \ref{dsm1}  continue with a Monte Carlo study to see that the asymptotic approximation with the normal distribution is adequate. All sections end with fitting the case study data to the model of the section. 

\subsection{Data: AOK HCD} \label{sec11}

Even though our aim is to sample from the German population of 2004, we are restricted to sampling from the 25 million members of   Germany's largest public health insurance company `Allgemeine Ortskrankenkasse' (AOK). This represents one-third of the population, presumably with, on average, slightly elevated disease rates, compared with other statutory health insurance funds or private health insurance funds \cite[see][]{bocken2008}. As we are only interested in a difference between two intensities of disease onset - with and without preceding stroke, we refrain from studying selection bias in that respect. The AOK's health claims data (HCD) include information about age, year of birth and date of exit (death or switch to another insurance company). From the insurance inventory on 01/01/2004, a simple sample of 250,000 people is drawn. We follow the health histories of those persons until the end of 2013. We exclude 4121 persons with implausible information over time on sex, birth year or region of living. In that form, the data are sufficient for the lifetime state model in Appendix \ref{lsm}.
The AOK HCD also contain information about outpatient and inpatient diagnoses for each insured person, with at least one day of insurance coverage, regardless of whether or not they sought medical treatment.
Recall, that we study here mainly the design effect that by sampling from the above population in 2004, people who died earlier an not be selected, i.e. are left-truncated. There is a second design effect which we, however, will not discuss in depth. We draw in 2004 only a sample of those persons at ages 50 years or older. 
That age is typically the earliest at which a stroke or dementia occurs. Of course, age 50 is not the earliest age at which a person may die and, additionally, left-truncated is a person who dies before age 50. We refrain from considering the second truncation reason, i.e. assume death before 50 impossible, for the sake of model simplicity and trade-off two effects. On the one hand, only $\approx$ 6.5\% of people die before age 50 in western civilisations.\footnote{https://www.ssa.gov/oact/STATS/table4c6.html\#fn1} Hence assuming that rate to be zero, will not distort the results by much. On the other hand, our data donor AOK, or more precisely its scientific research institute (WIdO), allowed sampling 250,000 people and without restricting to those over 50 years old would have resulted in $\approx$ 50\% younger\footnote{https://service.destatis.de/bevoelkerungspyramide/index.html} and those will mostly stay healthy over the 10 years of observation. Hence doing so would increase the standard errors by at least $\sqrt{2}$ - 1 $\approx$ 40\%. In the AOK HCD, all diagnoses are coded in the International Statistical Classification of Diseases and Related Health Problems (ICD), revision 10, issued by the WHO. Dementia was defined as having at least one of the following diagnoses coded by ICD-10: G30, G31.0, G31.82, G23.1, F00, F01, F02, F03, and F05.1. Note that by sampling at the beginning of 2004, a person with dementia diagnoses at that time (technically in that quarter or the next) may not indicate a dementia onset, but can be a prevalent case. We exclude those and observable remain {\em n} = 236,039 persons. For those, the mean follow-up time is 7.3 years, resulting in 1.7 million person-years at risk.  Some more descriptive cross-sectional statistics as of 2004, not necessarily needed in the models, are given in Table \ref{desstat}. 
\begin{table}[h]
	\centering
	\caption{Descriptive statistics for AOK insurants 50+ sampled in 2004, cross-section mid of 2004, except (longitudinally): mean follow-up time and person-years at risk }
	\label{desstat} \smallskip
	\renewcommand{\arraystretch}{0.7}
	\begin{tabular}{lrrlrr}
		\hline
		Variable	&		Number	&	\%	& Variable	&  Number & 	\% \\ \hline
		\underline{Age group}	&  & & \underline{Sex} &  & \\
		$[$50,55)	&	37,635	&	15.9 &	Men	&	101,779 & 43.1\\ 
		$[$55,60)	&	29,002	&	12.3 	& Women	&	134,260	&	56.9 \\ 
		$[$60,65)	&	36,419	&	15.4	&  \underline{Stroke} & & \\
		$[$65,70)	&	42,398	&	18.0	&		No	&	230,175	&	97.5 \\
		$[$70,75)	&	33,411	&	14.2	&		Yes	&	5864	&		2.5 \\
		$[$75,80)	&	27,062	&	11.5	&  \underline{Multi-morbid} & &  \\
		$[$80,85)	&	18,984	&	8.0	&	No &	144,816	& 61.4	 \\
		$[$85,90)	&	6318	&	2.7	&	Yes	& 91,223	&	38.6 	 \\
		$[$90,95)	&	4001	&	1.7	& & & \\
		$[$95,100)	&	744	&	0.3	&	Total & \multicolumn{2}{l}{236,039 persons} \\
		$[$100,105)	&	62	&	0.03  & Mean follow-up time	& \multicolumn{2}{l}{7.31 years}	 \\
		$[$105,110)	&	3	&	0.001 & Person-years at risk & \multicolumn{2}{l}{1,724,296 years} \\ \hline
	\end{tabular}
	\renewcommand{\arraystretch}{1}
\end{table}
Important longitudinal information for the models will be 34 thousand persons that will experience dementia onset up to 2013. Finally, in view of assessing the effect of a stroke (ICD-10: I63, I64) on dementia onset, due to the Markov property we will assume for all models, the information about a stroke for 5864 persons in 2004, more likely to be new than not, did not cause us to remove the persons. Additionally, 19,201 persons will experience a stroke up to 2013. The definition and relevance of multi-morbidity will be explained in Section \ref{dsm3}.

\subsection{Literature review}

Similar ideas of testing the effect of stroke on dementia are found in \cite{desmond2002}, \cite{ivan2004}, \cite{reitz2008}, \cite{savva2010}, \cite{kuzma2018}, \cite{kimlee2018} and \cite{hbid2020}. Let us compare our contribution broadly to the adjacent literature, distinguishing substantial and methodological similarities. Substantially, \cite{vieira} report dementia incidences, as do \cite{leys2005} after stroke. Death incidences, after stroke \citep{vandenBus} and with dementia \citep{garcia}, are of use for us because they constitute elements for one of our models and will enter the calibration of simulations.  Dementia prevalence is studied in \cite{jessen2018} and risk factors are presented in \cite{mangia2012}.
Community-based studies on the effect of stroke on dementia are \cite{ivan2004} and \cite{reitz2008}. Cerebrovascular processes are studied in more detail by \cite{huchen2017}. Common statistical risk factors to dementia and stroke are studied in \cite{pend2009}. With respect to the method, our work has considerable similarity with the study `Mortality of Diabetics in the County of Fyn' in \cite{And}, and \citet[][Section 4]{And0} in particular. However, our truncation model is slightly easier, our simple random sample of HCD is considerable larger than the data there, and we reduce the arguments to those necessary for our model. We make considerable use of \cite{flem1991}, even though the book does not cover left-truncation. \cite{testing-ho:2006} and \cite{A-likeliho:2009} apply a similar Markovian multi-state model, however to an economic question, and especially, need to reduce their population in order to circumvent left-truncation. Note that right-truncation needs another method \cite[see e.g.][and references therein]{Doe,weiswied2021,weisdoer2022}. Finally, \cite{weisetal2020} also analyse the same dataset and with dementia as endpoint, however, not using a multi-state model and with an emphasis on left-censoring.

\section{\sc Univariate analysis of health states} \label{dsm1}

The set of assumptions in the present section is to some extend educational, because the assumptions will be too restrictive for realistic results on the development of dementia. The results may not be compared to the epidemiological literature. Section \ref{dsm2} will present a first realistic  model, requiring more notation. Here we only explain major modelling decisions.  Following up on Section \ref{sec11}, let $t$ count the years after a person's 50$^{th}$ birthday and we continue to call $t$ `age'. The major methodological challenge is that people from our population of interest have died before 2004, the year when we started to observe. Their health histories have been deleted after their deaths, i.e. we cannot observe any of those histories, even if some of these people would belong to a simple sample from that population. Methodological arguments how to precede when drawing from the conditional population of alive people in 2004, are discussed in Appendix \ref{lsm} in a lifetime state model with death being both, the sole event of interest and the event of truncation. To augment the scope from mortality to morbidity also, let here $X_t$ indicate a person's disease state, $H$, $S^1$ or $D$, or $d$, at the age of $t$ (see Figures \ref{gesch} and \ref{MarkovProzessBild}). An healthy individual's status is set to $S^1$ at the time of the first stroke, and maintained (in the absence of further state transitions), also in case of subsequent strokes. For a person with dementia we write $X_t=D$, irrespective of whether a stroke has preceded dementia onset at that age or not. A stroke after dementia onset is not recorded at all, as it is not relevant for the assessment of a stroke effect on dementia. Now, with the - in comparison to mortality analysis - two additional states, $S^1$ and $D$, no transition into any other state than $d$ before 2004 prohibits observing the health history from 2004 on. Only death truncates a person. We will often refer the reader to Appendix \ref{lsm} for detailed analytical arguments and amend arguments in this section only when the disease state model differs methodologically from the lifetime state model. 

\subsection{Contribution of an observed person to inference} \label{modeldsm}

Roughly speaking, we aim at maximum likelihood inference. In case of a simple sample, each randomly drawn person contributes with its density to the likelihood, and the estimation criterion is the maximisation of the joint density as a function of the parameter, i.e. the likelihood. We will see that people not observed do not contribute to our criterion function and we now derive the contribution for each observed history. We first collect all possible state transitions in the index set 
\[
\mathcal{I}:=\{H S^1, S^1 D, H D, H d, S^1 d, D d \} \quad \text{(see Figure \ref{gesch})}.
\] 
Furthermore, universally for all persons, we do not follow a health history any further than $\tau$ years. The continuous-time history $\mathbf{X}=\{X_t, t\in [0,\tau]\}$, observed in full, in parts or not at all, defined on the probability spaces $(\Omega, \mathcal{F}, P_{\boldsymbol{\lambda}})$ represent either the population or one random draw from it.
We assume throughout the Markov property, so that the history is determined by the transition intensities $\lambda_{hj}(t):= \lim_{s\searrow 0} P_{\boldsymbol{\lambda}} (X_{t+s}=j \mid X_t=h)/s$. In this section, we model the population of Germany (at that time) as age-homogeneous, i.e. assume $\lambda_{hj}(t) \equiv \lambda_{hj}$. (The realistically age-inhomogeneous intensities follow in Section \ref{dsm2}.) By parameter we mean the vector $\boldsymbol{\lambda}:= (\lambda_{hj}, hj \in \mathcal{I})'$. We consider a simple random sample of size $n_{all}$ persons drawn from the population (see Figure \ref{data_selection}). The generalisation, as compared to Appendix \ref{lsm}, is theoretically less severe when we assume that all persons start in the same state, $X_0=H$, at age origin. Practically, from those not truncated by death, denoted as $n$ in Appendix \ref{case1}, $X_0 \in \{S^1,D\}$ is known to be rare, still exclusion is impossible because, e.g. for an observed person with $X_u=D$, $X_0$ is unknown. An option is to condition on the distribution of $X_0$, which leaves the criterion as a function of $\boldsymbol{\lambda}$ unchanged, if the distribution of $X_0$ does not depend on $\boldsymbol{\lambda}$. Sloppily, with $\mathcal{L}$ denoting the distribution, this is because we can decompose $\mathcal{L}_{\boldsymbol{\lambda}}(\mathbf{X})=\mathcal{L}_{\boldsymbol{\lambda}}(\mathbf{X}|X_0)\mathcal{L}(X_0)$. (In the case of $\mathcal{L}_{\boldsymbol{\lambda}}(X_0)$, efficiency gets lost.) We pursue the option for the AOK HCD. An additional benefit is notational simplicity, because we again observe those same $n<n_{all}$ persons (as in Appendix \ref{lsm}), with histories that occur -- completely or in part -- during our observation period between 2003 and 2014 (see Figure \ref{data_selection}). 

For each observed history, the `age-at-study-begin', $U$, is the time between the calendar dates of the 50$^{th}$ birthday and the study begin on 01/01/2004. 
In Appendix \ref{case1} we initially simplify to a non-random age-at-study-begin, $u$, and the age-at-death $T$ is reformulated as (jump-diffusion) process $N_T$. Here we generalise and reformulate one history $\mathbf{X}$ in several $N_{hj}(t) :=  \sum_{s \le t} \mathds{1}_{\{X_{s-} =h, X_s=j\}}$, the
processes `counting' the transition between states, up to age $t$, and $Y_h(t) := \mathds{1}_{\{X_{t-} =h\}}$, indicating residence in state $h$, at the age of $t$. The counting processes are collected in the vector $\mathbf{N}_X(t):=(N_{hj}(t), hj \in \mathcal{I})'$. 
Now, as usual, statistical statements about parameters are deferred from statements about the location (the `signal') of the random experiment. In order to define a location for a stochastic process, probabilities may be calculated on a filtration $\mathcal{N}_t:= \sigma\{\mathbf{N}_X(s), 0 \le s  \le t\}$. We may assume that $\mathbf{N}_X(t)$ is adapted to it, because we theoretically assume $X_0=H$. We assume for the `true', the population, parameter  $\boldsymbol{\lambda}_0$ (similar to \ref{pararaumeinfach} in Appendix \ref{lsm}):
\begin{enumerate}[label=\AnnBNumm]
	\item \label{parraummodb} It is $\lambda_{hj^0} \in \Lambda_{hj}:=[\varepsilon_{hj};1/\varepsilon_{hj}]$ for some small $\varepsilon_{hj} \in (0,1)$.
\end{enumerate}
The compensator of $\mathbf{N}_X$, the location concept here, has an intensity (intuitively the derivative of the compensator) with respect to $\mathcal{N}_t$ and $P_{\boldsymbol{\lambda}}$ of
$\boldsymbol{\alpha}(t):=(Y_h(t) \lambda_{hj}, hj \in \mathcal{I})'$. (When only $h$ appears, the first position of $hj$ it meant. Especially $h \neq d$, because death is absorbing.) Note that the notion of a `derivative' from real analysis, being with respect to the Lebesgue measure, needs a bit generalisation here.  
Starting with deterministic $u$, a person is not left-truncated in case of $A:=\{X_u \neq d \}$. Different to the lifetime state model is that the history up to $u$ is only known when $X_u=H$. If for instance $X_u=S^1$, the age-at-stroke is left-censored. This is an important incentive to start observation only at $u$, i.e. $\leftindex_u{\mathbf{N}}(t):=\mathbf{N}_X(t) - \mathbf{N}_X(t \land u)$, where $t \land u := \min(t,u)$. Due to the Markovian property it is adapted to the filtration $\leftindex_u{\mathcal{G}}_t  := \sigma\{\leftindex_u{\mathbf{N}}(s), u \le s \le t\}$.   
\begin{lemma}
	With respect to the probability measure  $P^{A}_{\boldsymbol{\lambda}}(F):= P_{\boldsymbol{\lambda}}(F \cap A)/P_{\boldsymbol{\lambda}}(A)$ for $F \in  \mathcal{F}$,
	the intensity of $\leftindex_u{\mathbf{N}}(t)$ is  
	$\leftindex_u{\boldsymbol{\alpha}}(t):=\mathds{1}_{\{u < t\}} \boldsymbol{\alpha}(t)$.
\end{lemma}

The proof is as in Section \ref{ltrcfixed}. 
Note that $P^A_{\boldsymbol{\lambda}}$ depends on the parameter $\boldsymbol{\lambda}$ and on $u$. With $\leftindex_u{Y}_h(t):= \mathds{1}_{\{u < t\}}Y_h(t)$, the coordinates of $\leftindex_u{\boldsymbol{\alpha}}$ are
$\leftindex_u{\alpha}_{hj}(t):= \leftindex_u{Y}(t) \lambda_{hj}$.
The observed left-truncated and right-censored counting process is $\leftindex_u{\mathbf{N}}^c(t) := \int_0^t C(s) d \leftindex_u{\mathbf{N}}(s)$,
with $C(t):= \mathds{1}_{\{t \le u+10\}}$ and $\leftindex_u{\mathbf{Y}}^c(t):=C(t) \leftindex_u{\mathbf{Y}}(t)$ 
\cite[compare][Example 1.4.2]{flem1991}. It has intensity
\begin{equation} \label{intesmodelbu}
	\leftindex_u{\boldsymbol{\alpha}}^c(t) := \mathds{1}_{\{t \le u+10\}} \leftindex_u{\boldsymbol{\alpha}}(t)=  \mathds{1}_{\{u < t \le u+10\}} \boldsymbol{\alpha}(t) 
\end{equation}
with respect to $P^A_{\boldsymbol{\lambda}}$ and observed filtration $\leftindex_u{\mathcal{F}}^c_t := \sigma\{\leftindex_u{\mathbf{N}}^c(s), u \le s \le t\}$. For the distinction be observable and unobservable filtrations see Section \ref{mortalran}.  As $\leftindex_u{\mathcal{F}}^c_t$ is a required self-exciting filtration, by Jacod's formula \cite[see][Formula (4.3)]{And0}, the contribution of a person (truncated or not) to the marginal likelihood and its (natural) logarithm are:
\begin{eqnarray} \label{likefixeeha}
	dP & = & \Prodi_{u < t \le u+10} \{ (1-\leftindex_u{\alpha}_{\cdot}^c(t)dt)^{1- d\leftindex_u{N}_{\cdot}^c(t) } \prod_{hj \in \mathcal{I}} (\leftindex_u{\alpha}_{hj}^c(t) dt)^{d\leftindex_u{N}_{hj}^c(t)} \} \nonumber \\
	& = & \left[\prod_{\stackrel{t\in [u,u+10], \exists hj \in \mathcal{I}}{\leftindex_u{N}^c_{hj}(t-) \not= \leftindex_u{N}^c_{hj}(t)}}\prod_{hj \in \mathcal{I}} \left(\leftindex_u{Y}^c_{h}(t)\lambda_{hj}\right)^{\Delta \leftindex_u{N}^c_{hj}(t)}\right] \nonumber \\
	& & \exp \left(-\sum_{hj \in \mathcal{I}} \int_u^{u+10} \leftindex_u{Y}^c_{h}(t)
	\lambda_{hj} dt\right) \nonumber \\
	\ln dP & = & \int_0^{\tau} \sum_{hj \in \mathcal{I}} \ln (\leftindex_u{Y}^c_h(t) \lambda_{hj}) d \leftindex_u{N}^c_{hj} (t)  - \sum_{hj \in \mathcal{I}} \int_0^{\tau} \leftindex_u{Y}^c_h(t) \lambda_{hj} dt 	
\end{eqnarray}
Note $\leftindex_u{\alpha}_{\cdot}^c(t) := \sum_{hj \in \mathcal{I}} \leftindex_u{\alpha}_{hj}^c(t)$ together with $\leftindex_u{N}_{\cdot}^c(t) := \sum_{hj \in \mathcal{I}} \leftindex_u{N}_{hj}^c(t)$. The product integral $\prodi$ is explained in Appendix \ref{ltrcfixed}. Essentially, the discrete approximation of the history $\mathbf{X}$ is a collection of random increments. The probability function (pf) of this collection can be a product of the increments' pf's. Decreasing the grid spacing defines an integral. The double-use of the integration symbol $dt$ in the first line is still different to the line above \eqref{lihoodexp} (in Appendix \ref{ltrcfixed}) because $Y(t)$ drops to zero after $T$, whereas $Y_h(t)$ is only one for a different state. The reason for the exponential function in the third line is explained shortly after \eqref{lihoodexp}. For the second equality, on the logarithmic scale, the logarithm of a product becomes a sum of logarithms and no new integration arises, the Stieltjes integration for discontinuous $g$ ($\int fdg=\sum f \Delta g$) suffices. Note that $\Delta \leftindex_u{N}^c_{hj}(t)= \leftindex_u{N}^c_{hj}(t) - \leftindex_u{N}^c_{hj}(t-)$ is only not zero if $\leftindex_u{N}^c_{hj}(t)$ jumps. These jumps are of height one. Further note that $\int_u^{u+10}$ can be replaced by $\int_0^{\tau}$, because $\leftindex_u{Y}^c_h(t)$ already accounts for the limits, and similarly, in the product, $[u,u+10]$ is accounted for in $\leftindex_u{\mathbf{N}}^c(t)$. 
Note that, because almost surely $\leftindex_u{N}^c_{hj} (u)=0$,
\begin{equation} \label{score1}
	\frac{\partial}{\partial \lambda_{hj}}  \ln dP =   \frac{\leftindex_u{N}^c_{hj} (\tau)}{\lambda_{hj}}  -  \int_0^{\tau} \leftindex_u{Y}^c_h(t) dt. 	
\end{equation}

A truncated person does not contribute to the marginal likelihood, as argued in detail with Formula \eqref{lihoodexp} in Appendix \ref{lsm}. As $\mathbf{X}$ is random, so must be the age-at-study-entry, $U$.
Similar to \ref{moeglich} in Appendix \ref{mortalran}, together with independent truncation, we impose as additional assumption, that not everybody is dead, prior to 2004: 
\begin{enumerate}[label=\AnnBNumm]
	\setcounter{enumi}{1}
	\item \label{posobsprobmodelb} $U$ and $\mathbf{X}$ are independent, it is $A:=\{ X_U \neq d\}$ and $\beta_{\boldsymbol{\lambda}_0}:= \tilde{P}_{\boldsymbol{\lambda}_0}(A) > 0$
\end{enumerate}
The additional information by stopping time $U$, i.e. for $(\mathbf{X},U)$, and at the same time the loss in information by truncation, is reflected by including $\leftindex_U{\mathbf{Y}}^c$ in the filtration $\leftindex_u{\mathcal{F}}^c_t$, $\leftindex_U{\mathcal{F}}^c_t := \sigma\{\leftindex_U{\mathbf{N}}^c(s), \leftindex_U{\mathbf{Y}}^c(s), u \le s \le t\}$. Consult Appendix \ref{mortalran} to see that, similar to non-random truncation \eqref{likefixeeha}, conditional on the last two coordinates,  by Jacod's formula the logarithmic density of $(\mathbf{X},U, \mathds{1}_A)$ up to $\tau$ is 
\begin{equation} \label{likeraneha}
	\ln \leftindex_U{L}_{\tau}^c(\mathbf{X},U|\boldsymbol{\lambda})  =  \int_0^{\tau} \sum_{hj \in \mathcal{I}} \ln (\leftindex_U{Y}^c_h(t) \lambda_{hj}) d \leftindex_U{N}^c_{hj} (t)  
	- \sum_{hj \in \mathcal{I}} \int_0^{\tau} \leftindex_U{Y}^c_h(t) \lambda_{hj} dt, 
\end{equation} 
where $U$ replaces $u$ in the definitions of $\leftindex_u{N}^c_{hj} (t)$, $\leftindex_u{Y}^c_h(t)$ and $\leftindex_u{\boldsymbol{\alpha}}^c(t)$ of \eqref{intesmodelbu}. The expression uses the Doob-Meyer decomposition of $N_{hj}(t)$, stacked to $\mathbf{N}_X$. Occasionally,  we will denote the second term in \eqref{likeraneha}, and in corresponding decompositions for more advanced models, the subtrahend, as `$Y$-term'.
The first term, the minuend, will be denoted as the `$N$-term', because $Y$ will vanish after taking derivatives, essentially due to $\frac{d}{dx} \int=\int \frac{d}{dx}$ and $\ln (a x)=\ln (a)+ \ln(x)$.
The observed left-truncated and right-censored versions thereof are 
\[
\leftindex_U{\mathbf{N}}^c(t) := \int_0^t C(s) d \leftindex_U{\mathbf{N}}(s) \quad 
\text{and} \quad  \leftindex_U{Y}^c_h(t):= C(t) \mathds{1}_{\{U < t\}}Y_h(t),
\]
with $C(t)$ being one, as long as the person is not censored, i.e. for $t \le U+10$ and $\leftindex_U{\mathbf{N}}(t):=\mathbf{N}_X(t) - \mathbf{N}_X(t \land U)$. That the contribution of a truncated person to the conditional likelihood is one, is argued in Appendix \ref{mortalran}, Formula \eqref{trunccenslihoodexp}. 


\subsection{Point estimates and their standard errors}

\begin{figure}[t]
	\begin{center}
		\begin{picture}(280, 180)
			
			
			\put( 0.0, 15.0){\thinlines\vector(1, 0){290.0}}
			\put(34.0, 14.0){\thinlines\line(0, 1){2.0}}

			\put( 34.0, 5.0){\footnotesize{$0$}}
			\put(274.0, 5.0){\footnotesize{$t$}}
			
			\put(34.0, 65.0){\thinlines\line(1, 0){220.0}}
			\put(34.0, 95.0){\thinlines\line(1, 0){220.0}}
			\put(34.0, 125.0){\thinlines\line(1, 0){220.0}}
			\put(34.0, 155.0){\thinlines\line(1, 0){220.0}}
			
			
			\multiput(269.0,  65.0)(-3,0){25}{\thinlines\line(-1, 0){1.5}}
			\multiput(269.0,  95.0)(-3,0){25}{\thinlines\line(-1, 0){1.5}}
			\multiput(269.0, 125.0)(-3,0){25}{\thinlines\line(-1, 0){1.5}}
			\multiput(269.0, 155.0)(-3,0){25}{\thinlines\line(-1, 0){1.5}}
			
			\multiput(269.0, 65.0)(0,30){4}{\thinlines\vector(1, 0){5.0}}
			
			\put( 34.0, 64.0){\thinlines\line(0, 1){2.0}}
			\put( 66.0, 64.0){\thinlines\line(0, 1){2.0}}
			\put( 46.0, 64.0){\thinlines\line(0, 1){2.0}}

			\put( 34.0, 94.0){\thinlines\line(0, 1){2.0}}
			\put( 73.0, 94.0){\thinlines\line(0, 1){2.0}}
			\put( 153.0, 94.0){\thinlines\line(0, 1){2.0}}
			\put(103.0, 94.0){\thinlines\line(0, 1){2.0}}

			\put( 34.0, 124.0){\thinlines\line(0, 1){2.0}}
			\put( 76.0, 124.0){\thinlines\line(0, 1){2.0}}

			\put( 34.0, 154.0){\thinlines\line(0, 1){2.0}}
			\put(232.0, 154.0){\thinlines\line(0, 1){2.0}}
			
			\put( 36.0, 68.0){\tiny{$H$}}
			\put( 54.0, 68.0){\tiny{$S^1$}}
			\put(179.0, 68.0){\tiny{$D$}}
			
			\put( 55.0, 98.0){\tiny{$H$}}
			\put( 86.0, 98.0){\tiny{$S^1$}}
			\put( 126.0, 98.0){\tiny{$D$}}
			\put(182.0, 98.0){\tiny{$d$}}

			\put( 53.0, 128.0){\tiny{$H$}}
			\put(161.0, 128.0){\tiny{$D$}}
			
			\put( 121.0, 158.0){\tiny{$H$}}
			\put(241.0, 158.0){\tiny{$D$}}
			
			\put(18, 65.0){\footnotesize{$X_t^1$}}
			\put(18, 95.0){\footnotesize{$X_t^2$}}
			\put(18, 125.0){\footnotesize{$X_t^3$}}
			\put(18, 155.0){\footnotesize{$X_t^4$}}
			
			\put( 34.0, 15.0){\textcolor[rgb]{0.6,0.6,0.6}{\thicklines\line(1, 0){32.0}}}
			\put(66,15){\textcolor[rgb]{0.6,0.6,0.6}{\circle{3.5}}}
			\put(66,35){\textcolor[rgb]{0.6,0.6,0.6}{\circle*{3.5}}}
			\put( 66.0, 35.0){\textcolor[rgb]{0.6,0.6,0.6}{\thicklines\line(1, 0){37.0}}}
			\put(103,35){\textcolor[rgb]{0.6,0.6,0.6}{\circle{3.5}}}
			\put(103,55){\textcolor[rgb]{0.6,0.6,0.6}{\circle*{3.5}}}
			\put(103.0, 55.0){\textcolor[rgb]{0.6,0.6,0.6}{\thicklines\line(1, 0){160.0}}}
			
			\put(263.0, 55.0){\textcolor[rgb]{0.6,0.6,0.6}{\thicklines\vector(1, 0){5.0}}}
			
			\put(67.0, 40.0){\tiny{\textcolor[rgb]{0.6,0.6,0.6}{$N_{S^1D\bullet}(t)$}}}

			\put( 34.0, 17.0){\thicklines\line(1, 0){68.0}}
			\put(103,17){\circle{3.5}}
			\put(103,33){\circle*{3.5}}
			\put(104.0, 33.0){\thicklines\line(1, 0){157.0}}
			\put(263.0, 33.0){\thicklines\vector(1, 0){5.0}}

			\put(167.0, 41.0){\footnotesize{$\leftindex_U{N}^c_{S^1D\bullet} (t)$}}
			
			
			\multiput(66.0, 154.0)(0,-3){47}{\thinlines\line(0, -1){1.5}}
			\multiput(103.0, 154.0)(0,-3){47}{\thinlines\line(0, -1){1.5}}
			
			
			\put(154.0, 61.0){\thicklines\line(0, 1){7.0}}
			\put(152.0, 71.0){\footnotesize{$U_1$}}
			
			\put(44.0, 91.0){\thicklines\line(0, 1){7.0}}
			\put(42.0, 101.0){\footnotesize{$U_2$}}
			
			\put(144.0, 121.0){\thicklines\line(0, 1){7.0}}
			\put(142.0, 131.0){\footnotesize{$U_3$}}
			
			\put(84.0, 151.0){\thicklines\line(0, 1){7.0}}
			\put(82.0, 161.0){\footnotesize{$U_4$}}
			
		\end{picture}
		\caption{Four health histories (end of states is indicated by short vertical line, left-truncation events at $U_i$ are indicated by long vertical lines), unobservable $N_{S^1D\bullet}(t):=\sum_{i=1}^n N_{S^1D i}(t)$ and observed $\leftindex_U{N}^c_{S^1D\bullet} (t)$ (offset between $N_{S^1D\bullet}$ and $\leftindex_U{N}^c_{S^1D\bullet}$ for clarity)} \label{MarkovProzessBild}
	\end{center}
\end{figure}
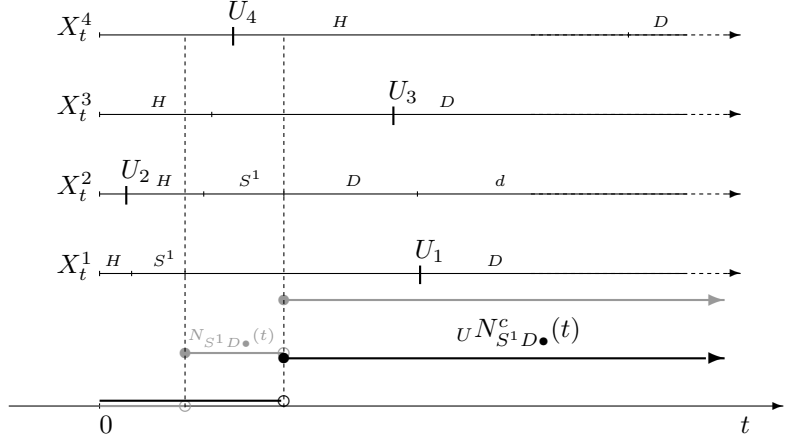

As in Appendix \ref{appsec4}, $i$ denotes the person in $\mathbf{X}^i$, $U_i$, $\leftindex_{U}{N}^c_{hji}$ and 
$\leftindex_U{Y}^c_{hi}$ and still $n=\sum_{i=1}^{n_{all}} \mathds{1}_{A_i}$ with $A_i$ as in \ref{posobsprobmodelb}. The truncated persons without contribution to the conditional likelihood are sorted to the end of the unobserved sample, a convention already in \cite{heckman1976}. All others contribute with \eqref{likeraneha} to  
\begin{equation} \label{loglhomo}
	\ln \leftindex_U{L}_{\tau}^c(data;\boldsymbol{\lambda})  =  \sum_{i=1}^n  \ln \leftindex_U{L}_{\tau}^c(\mathbf{X}^i,U_i|\boldsymbol{\lambda}). 
\end{equation}
This requires $U_i$ to be random as explained in Appendix \ref{appsec4}. With $\leftindex_U{N}^c_{hj\bullet}(t):= \sum_{i=1}^n \leftindex_{U}{N}^c_{hji}(t)$ and 
$\leftindex_U{Y}^c_{h\bullet}(t):=\sum_{i=1}^{n}\leftindex_U{Y}^c_{hi}(t)$ (sketched in Figure \ref{MarkovProzessBild} for $n_{all}=4$)
the unique root of the derivatives of \eqref{loglhomo} - and hence the point estimates - are, by \eqref{score1} simply 
\begin{equation} \label{ehaest}
	\hat{\lambda}_{hj}  =  \frac{\leftindex_U{N}^c_{hj\bullet}(\tau)}{\int_0^{\tau} \leftindex_U{Y}^c_{h\bullet}(t) dt},
\end{equation}
One can avoid integration in the denominator in \eqref{ehaest}. Of the interesting states for $h$, $H$ and $S^1$, rewrite e.g. for $h=H$:
\begin{multline*}
	\int_0^{\tau} \leftindex_U{Y}^c_{H\bullet}(t) dt  
	=  \sum_{i=1}^n \int_{U_i}^{U_i + 10} \mathds{1}_{\{X^i_{t-} =H\}} dt \\
	=  \sum_{i=1}^n \mathds{1}_{\{X^i_{U_i} = H\}} \left( 10 \, \mathds{1}_{\{X^i_{U_i+10} = H\}} +  \mathds{1}_{\{X^i_{U_i+10} \in \{S^1,D\}\}} (\sup_t\{X^i_t=H\} - U_i) \right)
\end{multline*}

Note that similar to \eqref{111}, by using the simple sample assumption, among those who survive $U_i$ (i.e. 2003), the portion in the study period at age $t$ in state $h$ is asymptotically the same in the observed sample and in the entire population. By the LLN, for fixed $t$, 
\begin{equation} \label{mittel}
	\frac{\leftindex_U{Y}^c_{h\bullet}(t)}{n}\stackrel{P}{\longrightarrow}m^A_h(t):=\tilde{P}_{\boldsymbol{\lambda}_0}(X_t=h|A).
\end{equation}
The latter will typically be positive, but for our parametric model, we only need to assume (compare \ref{varpos2}): 
\begin{enumerate}[label=\AnnBNumm]
	\setcounter{enumi}{2}
	\item \label{varpos} $\int_0^{\tau} \sum_{h \in \{H, S^1, D\}} m^A_h(t)dt > 0$
\end{enumerate}

By verifying regularity conditions, we arrive at the (joint) asymptotic distribution of the estimators $\hat{\lambda}_{HD}$ and $\hat{\lambda}_{S^1D}$  by standard results on martingales. It depends on $m_h^A(t)$, the conditional prevalence of state $h$ at age $t$ in the population, and  $\beta_{\boldsymbol{\lambda}_0}$, the probablity of a person from the sample to be observed, i.e. not to be truncated.  

\begin{theorem}\label{normmodelb} Under Conditions \ref{parraummodb}-\ref{varpos} and $m^A_h(t)$ defined in \eqref{mittel} it is $\hat{\boldsymbol{\lambda}}$, composed of \eqref{ehaest}, consistent and
	$\sqrt{n_{all}}(\hat{\boldsymbol{\lambda}} - \boldsymbol{\lambda}_0) \stackrel{\mathcal{D}}{\longrightarrow} \mathcal{N}(0,\boldsymbol{\Sigma}^{-1}(\boldsymbol{\lambda}_0))$ with
	diagonal matrix $\boldsymbol{\Sigma}(\boldsymbol{\lambda}_0)$ of diagonal elements 
	\[ 
	\sigma_{hj,hj}(\boldsymbol{\lambda}_0):= \beta_{\boldsymbol{\lambda}_0}  \int_0^{\tau} \sum_{hj \in \mathcal{I}} \frac{ m^A_h(t) }{\lambda_{hj^0}}dt, \quad \text{for} \quad hj \in \mathcal{I}. 
	\]
\end{theorem}

Roughly speaking, the arguments of the proof, given in Appendix \ref{appb}, are similar to the case of the univariate parameter space in Appendix \ref{appsec4}. Luckily, the multivariate parameter space here results in a diagonal matrix of asymptotic variances, and positive definiteness follows from the positivity of the diagonal elements.

It remains to consistently estimate $\boldsymbol{\Sigma}(\lambda_0)$, in order to construct a confidence interval for the difference $\hat{\lambda}_{S^1D^0}-\hat{\lambda}_{HD^0}$ with the standard error. This then allows a Wald-type test for the effect of a stroke $S^1$ on the intensity of dementia onset for the AOK HCD in Section \ref{res2}. 
By Theorem \ref{normmodelb},
$Var(\hat{\boldsymbol{\lambda}}) = Var (\sqrt{n_{all}}\hat{\boldsymbol{\lambda}})/n_{all}\stackrel{\cdot}{=}\boldsymbol{\Sigma}^{-1}(\boldsymbol{\lambda}_0)/n_{all}$, so that, for estimating the asymptotic variance in Theorem \ref{normmodelb}, define $- \mathcal{J}_{\tau}(\boldsymbol{\lambda}_0)$ as
\begin{equation*}
	diag \left(\frac{\partial^2}{\partial \lambda_{hj}^2} \ln \leftindex_U{L}_{\tau}^c(data|\boldsymbol{\lambda})|_{\boldsymbol{\lambda}=\boldsymbol{\lambda}_0} ; hj \in \mathcal{I}\right)
	=  - diag \left(\frac{\leftindex_U{N}^c_{hj\bullet} (\tau)}{\lambda_{hj^0}^2}; hj \in \mathcal{I}\right). 
\end{equation*}
Now, as $\mathcal{J}_{\tau}(\boldsymbol{\lambda}_0)/n_{all} \stackrel{n_{all} \to \infty, P}{\longrightarrow}  \boldsymbol{\Sigma}(\lambda_0)$ \cite[see][Formula (6.1.11)]{And}, it is
\begin{equation}\label{ehastand}
	\widehat{Var(\hat{\lambda}_{hj})} =	 \mathcal{J}_{\tau}^{-1}(\hat{\boldsymbol{\lambda}})_{hj,hj}  =   \frac{\leftindex_U{N}^c_{hj\bullet} (\tau)}{\left(\int_0^{\tau} \leftindex_U{Y}^c_{h \bullet}(t) dt\right)^2}, 
\end{equation}
due to 
$\mathcal{J}_{\tau}^{-1}(\boldsymbol{\lambda}_0)_{hj,hj} = \lambda_{hj^0}^2/\leftindex_U{N}^c_{hj\bullet}(\tau)$, \eqref{ehaest} and the CMT.  The standard error of $\hat{\lambda}_{hj}$ is the square root thereof. Note that even though {\em n}$_{all}$, and with it the asymptotic variance - as component of $\boldsymbol{\Sigma}(\lambda_0)$ -, is not observable, the standard errors are indeed observed.    


\subsection{Simulation of finite sample properties} \label{finpropdsm}

Here we conduct a Monte Carlo simulation, primarily to visualize the asymptotic results on consistency, measured in (root) mean squared error, and on normality for small sample sizes, as indicated by Theorem \ref{normmodelb}. Especially we will find that the asymptotic approximation is rather precise for our statements on basis of the AOK HCD. Appendix \ref{simmort} does alike for the lifetime state model. We refrain from indicating the true parameter by the sub/superscript and drop $0$ in this section. We arrange $\boldsymbol{\lambda}$ as generator (left side):
\begin{equation} \label{qsim}
	Q:=\begin{pmatrix}
		-\lambda_{H \cdot} & \lambda_{H S^1} & \lambda_{H D} & \lambda_{H d} \\
		0 & -\lambda_{S^1 \cdot} & \lambda_{S^1 D} & \lambda_{S^1 d} \\
		0 & 0 & - \lambda_{D d} & \lambda_{D d}\\
		0 & 0 & 0 & 0
	\end{pmatrix} = \begin{pmatrix}
		-0.08\bar{6} & 1/30 & 0.02 & 1/30 \\
		0 & -0.17 & 0.07 & 1/10 \\
		0 & 0 & - 0.1 & 1/10 \\
		0 & 0 & 0 & 0
	\end{pmatrix}
\end{equation}
Here, the small dot signals summation over the respective index $\sum_{j}$. 

\subsubsection{Algorithm for simulating a history}

We simulate a disease state history with the description in \cite{Alb}. As discussed in Section \ref{modeldsm}, we assume $X_0=H$, so that $X_t \equiv H$ on $[0,T_1)$ with $T_1$ having the cumulative hazard function $A_{H\cdot}(t)= \lambda_{H\cdot} t$ (i.e. $T_1 \sim Exp(\lambda_{H \cdot})$). Then, in $t=T_1$, $\mathbf{X}$ migrates from $H$ to $j \in \{S^1, D, d\}$  with $p_{Hj}=\tilde{P}_{\boldsymbol{\lambda}}(X_{T_1}=j| X_{T_1-}=H)$, with $p_{Hj}= \lambda_{Hj} / \lambda_{H \cdot}$. Finally, (if $X_{T_1} \neq d$), $X_t \equiv j$ on $[T_1,T_2)$ with $T_2$ (and $j \in \{S^1,D\}$) having cumulative hazard function $A_{j\cdot}(t+T_1)- A_{j\cdot}(T_1)=  \lambda_{j\cdot} t$ (i.e. $T_2 \sim Exp(\lambda_{j \cdot})$). In $t=T_2$, $\mathbf{X}$ migrates from $j$ to $k \in \{D, d\}$  with $p_{jk}$.  Then (if $X_{T_2} \neq d$), $X_t \equiv D$ on $[T_2,T_3)$ with $T_3$ having cumulative hazard function $A_{Dd}(t+T_2)- A_{Dd}(T_2)= - \lambda_{Dd} t$ (i.e. $T_3 \sim Exp(\lambda_{Dd})$).   

\subsubsection{Section of true parameter, sample size and birth distribution} \label{sec232}

For a true parameter $\boldsymbol{\lambda}$ in a realistic region of the parameter space, intensities from the literature are reconciled with results the AOK HCD, anticipating Section \ref{res2}. The theoretical relation between incidences and intensities is given by $P(t)=e^{tQ}$ \cite[see e.g.][Formula (2)]{testing-ho:2006}, where $P(t)$ denotes the matrix of $t$-year probabilities $P(X_t=j| X_0=h)$. Similar to Section \ref{simmort}, the approximation $e^Q \approx I + Q$ (for $Q$ `small') allows to simply replace the one-year incidences for the intensities. 
For dementia onset after a stroke \cite{leys2005} find a one-year incidence of 7\%.  The value $\lambda_{S^1 D}$ = 0.07 will be confirmed for the AOK HCD in Section \ref{res2} (Table \ref{stathomo}). For dementia onset without a stroke, the AOK HCD result in $\lambda_{H D}$ = 0.02. Similarly, \cite{vieira} collect, but independent of whether a stroke preceded, one-year incidences of 0.008, 0.001 and 0.002 (dependent on the country and age range) for individuals below age 65. The AOK HCD value of 0.02 is larger, but aims at high ages as well and we stick to 0.02. 

There are other parameters, necessary for the simulation, but not of primarily of interest for the main question and will thus not be reported in  Section \ref{res2}. However, we can still estimate them from the AOK HCD using \eqref{ehastand} and compare with the literature. To start with, $\hat{\lambda}_{S^1 d}$ = 0.07; however, \cite{vandenBus} find for Germany, also from some other HCD, that 17\% of people die within one year after a stroke. We consider only the first stroke, which explains to some extend the smaller value. We use the value in between $0.1$. From the AOK HCD, $\hat{\lambda}_{D d}$ of 9\%, whereas \cite{garcia} find for Sweden from registry data that 11\% die each year with dementia. Of course, conceptionally, those that had died from other causes would need to be excluded, but we use as value in between 0.1. From AOK HCD, $\hat{\lambda}_{H S^1}$ = 0.02, whereas \cite{zhang} find incidences above 0.01 only for French people above age 80 and for Italian and British above the age of 75. We even opt to increase $\lambda_{H S^1}$ slightly further to 1/30 $\approx$ 0.03. For the death intensity without stroke or dementia $\lambda_{H d}$, we did not find a relevant study. Our general death hazard from Appendix \ref{simmort} is $\lambda$ = 0.02 and we increase slightly to 1/30 $\approx$ 0.03. Our final choice is collected in \eqref{qsim} (right-hand side).

As sample sizes we let {\em n}$_{all}$ vary from one to five, ten, 20 (and later 100) thousand people. All are below the sample size latent to the AOK HCD. However, we will see convergence to kick in, so that more computational burden is unnecessary. For the distribution of the age-at-study-begin $U$, we follow \cite{weiswied2021} and assume the distribution of $U$ to be uniform.

The longer the birth period, the more people are left-truncated, our population is born within 54 years (see Figure \ref{data_selection}). However, to start with, we only use 30 years, i.e. $U \sim U[0,30]$. Combined with \eqref{qsim}, on average, 48.7\% in the simulated samples are unobserved due to left-truncation. 

\subsubsection{Interpretation}

The number of simulation replications is 10,000. The simulation results in Table \ref{simbias} (top) confirm consistency of $\hat{\boldsymbol{\lambda}}$. Especially the root mean squared error drops, as a function of the sample size. The simulation averages of $\hat{\lambda}_{HD} - \lambda_{HD}$ (and similarly for transitions $S^1D$) reveal a generally small bias.  
\begin{table}[ht]
	\centering
	\caption{Simulated bias (Bias) and root means squared error (rMSE) for \eqref{ehaest} (top panel) and \eqref{ehastand} (bottom panel) and actual level for  confidence interval at the nominal level of 95\% for $\lambda_{HD}$ and $\lambda_{S^1D}$ (top panel) \eqref{ehastand} ({\em n} $\approx$ 0.513 $\times$ {\em n}$_{all}$, $\boldsymbol{\lambda}$ of \eqref{qsim})} \label{simbias}
	\begin{tabular}{rrrrrrr}
		\hline\noalign{\smallskip}
		\multicolumn{7}{c}{Estimator \eqref{ehaest} of $\boldsymbol{\lambda}$ (only transitions $HD$ and $S^1D$)} \\ \hline
		& \multicolumn{2}{c} {\textbf{Bias} ($\times$ 10$^2$)} & \multicolumn{2}{c} {\textbf{rMSE} ($\times$ 10$^2$)}  & \multicolumn{2}{c} {\textbf{act. level} (in \%)} \\
		{\em n}$_{all}$	& $\lambda_{HD}$ & $\lambda_{S^1D}$ &  $\lambda_{HD}$ & $\lambda_{S^1D}$ & $\lambda_{HD}$ & $\lambda_{S^1D}$ \\ 	\hline
		1000 & -0.0008 & 0.0352 & 0.3029 & 1.1435 & 94.39\% & 94.88\% \\ 
		5000 & 0.0003 & 0.0075 & 0.1345 & 0.5124 & 94.51\% & 95.20\% \\ 
		10,000 & -0.0009 & 0.0055 & 0.0947 & 0.3630 & 95.16\% & 94.82\% \\ 
		20,000 & -0.0005 & -0.0013 & 0.0663 & 0.2562 & 95.20\% & 95.01\% \\ \hline
		&  \multicolumn{5}{c}{Estimator \eqref{ehastand} of $Var(\hat{\boldsymbol{\lambda}})$ (only transitions$HD$ and $S^1D$)} &  \\ \hline
		&	& \multicolumn{2}{c} {\textbf{Bias} ($\times$ 10$^4$)}  & \multicolumn{2}{c} {\textbf{rMSE} ($\times$ 10$^4$)} &   \\
		&  {\em n}$_{all}$	& $Var(\hat{\lambda}_{HD})$ & $Var(\hat{\lambda}_{S^1D})$ &  $Var(\hat{\lambda}_{HD})$ & $Var(\hat{\lambda}_{S^1D})$ &  \\ 	\hline
		&   1000 & -0.0015 & 0.0415 & 0.0156 & 0.3022 &    \\ 
		& 5000 & -0.0001  & 0.0025  & 0.0014  & 0.0262 &     \\ 
		&  10,000 & 0.0000  & 0.0005  & 0.0005  & 0.0092 &    \\ 
		&  20,000 & 0.0001  & 0.0004  &  0.0002  & 0.0033 &   \\ 
		\hline
	\end{tabular}
\end{table}
The standard error \eqref{ehastand} can also be suspected to be consistent (see Table \ref{simbias}, bottom), without a formal proof in the above section. Simulations show similar behaviour for all other $\hat{\lambda}_{hj}$ (and their standard errors). The actual level of the confidence interval is close to the nominal. 

\begin{figure}[t]
	\centering
	
	\includegraphics[scale=0.251]{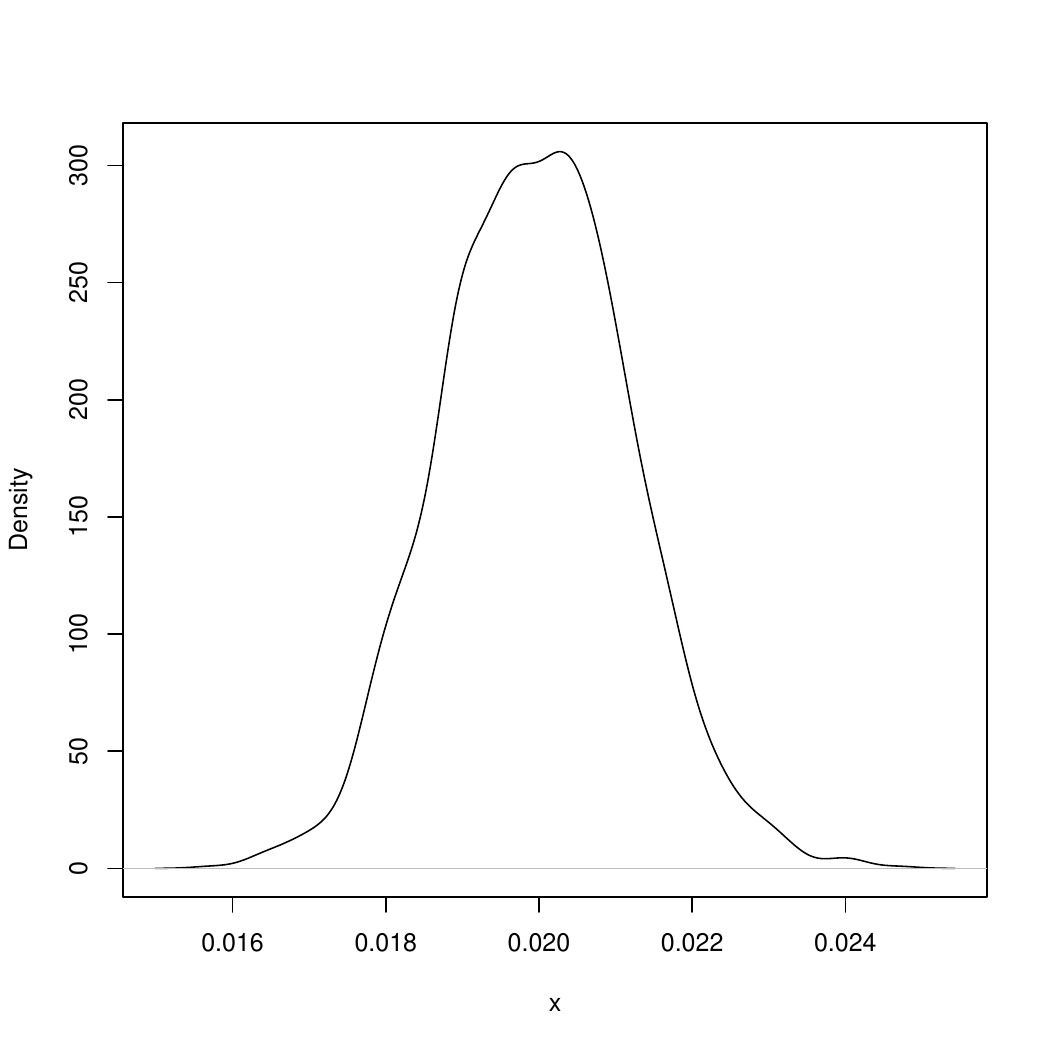}
	\includegraphics[scale=0.251]{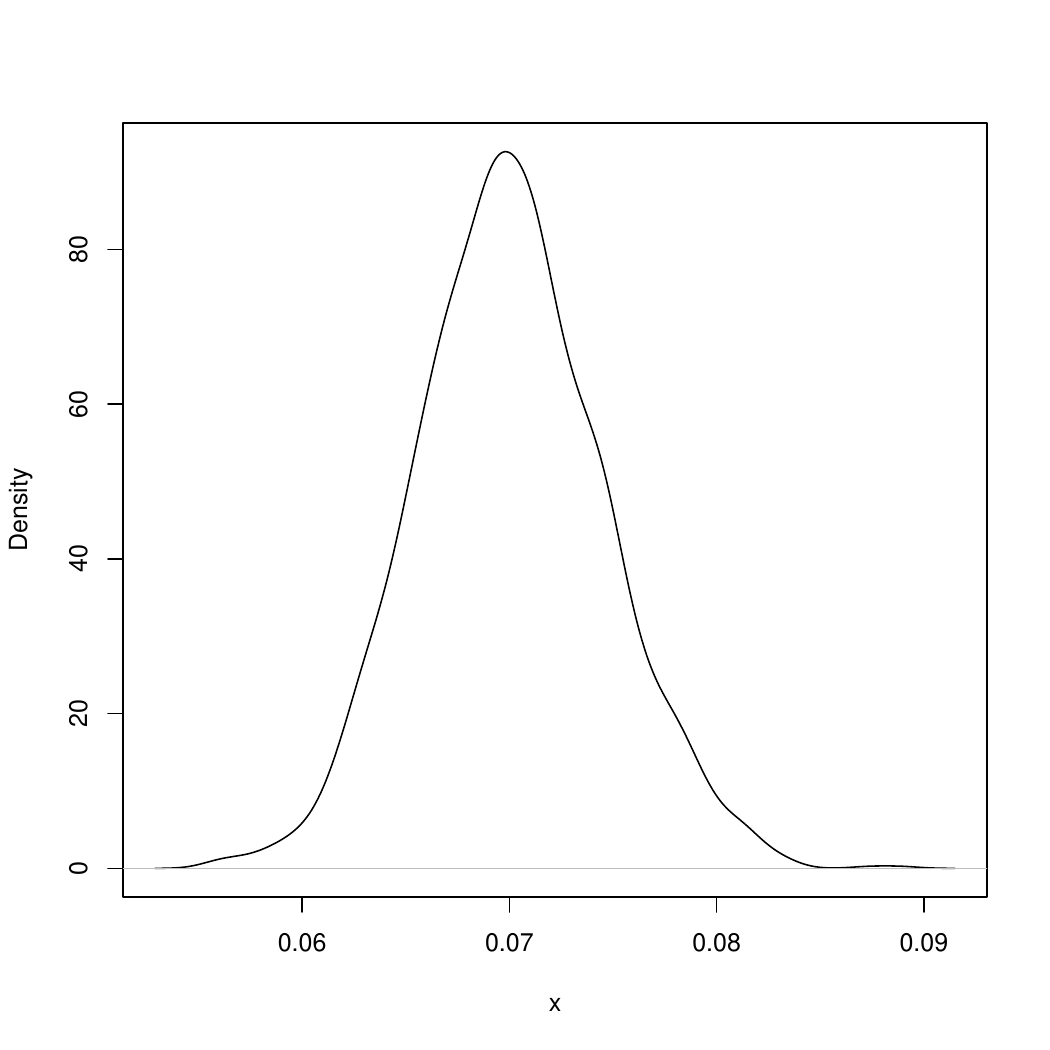}	
	\includegraphics[scale=0.251]{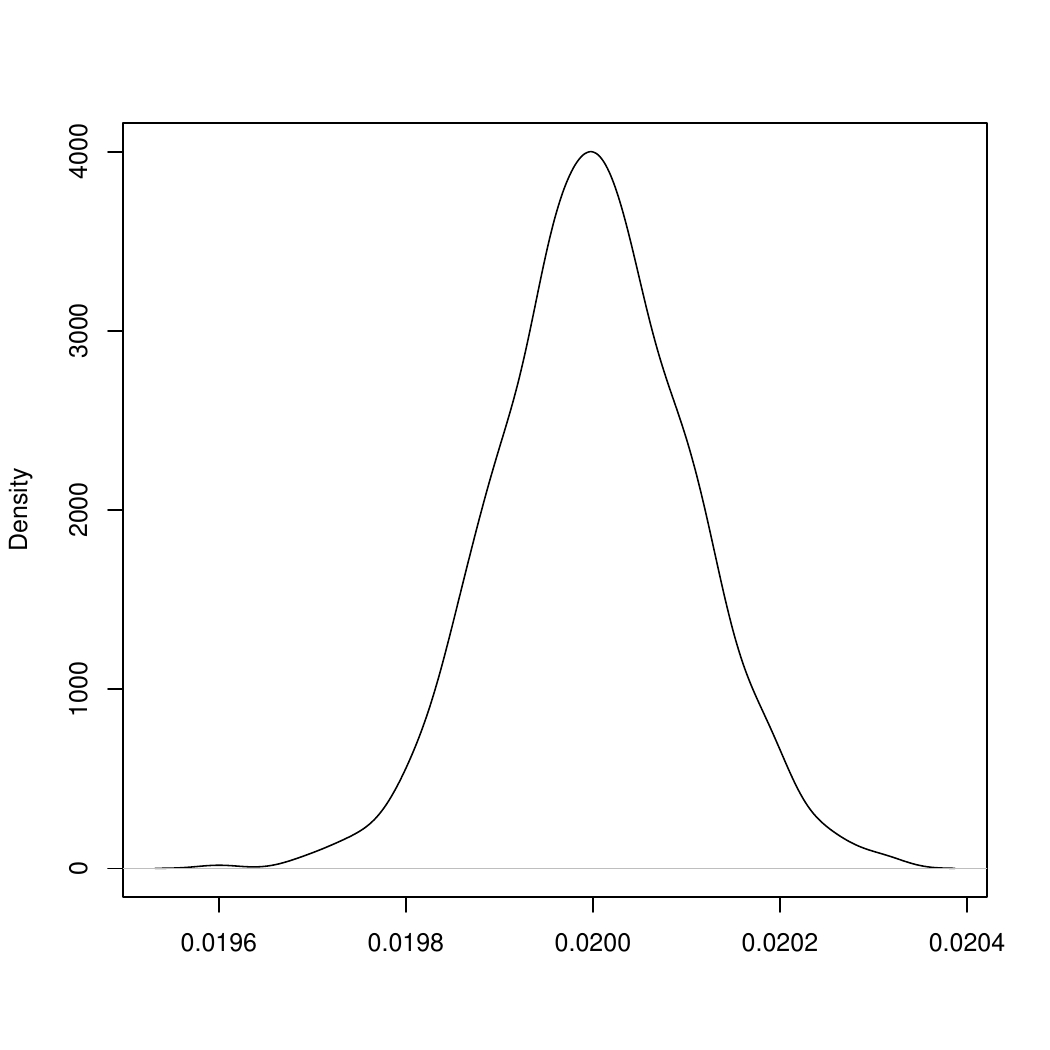}
	\caption{Distribution of estimators \eqref{ehaest} of dementia intensities $\lambda_{HD}$ = 0.02 and  $\lambda_{S^1D}$ = 0.07 (left/middle, $\approx$ 80\% truncated units) and of estimator \eqref{expest} for death hazard $\lambda_0$ = 0.02 (right, 40,000 uncensored observations), kernel smoothed from 2000 simulated samples with each {\em n} = 250,000 observations (Explanation of panels and symbols is distributed over larger parts of text.)}
	\label{simmorb}
\end{figure}

We now further approach the our population of the birth cohorts 1900 to 1953, i.e. now $U \sim  U[0,54]$, and use again \eqref{qsim}. We use {\em n}$_{all}$ = 100,000, being still below the sample size behind the AOK HCD, and run (only) 2000 simulation loops now.
The left and middle panel of Figure \ref{simmorb} confirm the asymptotic normality of $\hat{\lambda}_{HD}$ and $\hat{\lambda}_{S^1D}$ (of \eqref{ehaest}) stated in Theorem \ref{normmodelb}. The theorem also states asymptotic independence of the two estimators, which will be important when subsequently deriving a confidence interval for the difference. The simulated correlation $Cor(\hat{\lambda}_{HD}, \hat{\lambda}_{S^1D})$ = -0.02 confirms the independence.  


\subsection{Results for AOK HCD} \label{res2}

As population, we consider the 76 million Germans born between 01/01/1900 and 31/12/1954 (see Figure \ref{data_selection}). The data, i.e. the truncated sample, was described in Section \ref{sec11}. Recall that in the disease history $\mathbf{X}$, dementia at the age of $t$ is coded as $X_t=D$ (see beginning of Section \ref{dsm1}). We only remind here on the number of observations {\em n} = 236,039, and on  the maximally observed timespan $\tau$ := 54+10 = 64 years (after a person's 50$^{th}$ birthday). By doing so, the $n$ persons are at most followed until the age of 114 (see Figure \ref{c_figure}). The least possibly observed lifetime is just above 50 years, for a person turning 50 shortly before 01/01/2004 and dying shortly after that.  Preliminary results for the lifetime state model are in  Appendix \ref{case1}, where the hazard rate of the univariate lifetime has been estimated, and we expand our perspective now to the history of vascular diseases. We start from the logarithmic conditional likelihood \eqref{loglhomo} for the model introduced during Section \ref{modeldsm}. In the disease history model pursued here, in contrast to to `Mortality of Diabetics in the County of Fyn' \cite[see][]{And}, the age-at-study-begin, $U$, is linked to the age-at-study-end by $U$ + 10 \cite[compare][Examples III.3.6, IV.1.7 and VI.1.4]{And}.
As in \cite{hbid2020}, specifically, we compare dementia onset without a preceding stroke, $\lambda_{HD}$, to onset after a stroke, $\lambda_{S^1D}$. The data cover the information of 34,012 people with onset of dementia in the monitoring period, split into 6275 after a stroke and 27,737 not following a stroke (see Table \ref{stathomo}). Already with a stroke until 2004, 5864 persons (see Table \ref{desstat}) must be combined with 19,201 persons with newly diagnosed strokes between the forth quarter 2004 and the end of 2013.
\begin{table}
	\centering
	\caption{Statistics, point estimates \eqref{ehaest} and standard errors (SE, \eqref{ehastand}) for age-homogeneous model}
	\label{stathomo} \smallskip
	{\footnotesize
		\begin{tabular}{cccc}
			\hline
			\noalign{\smallskip}
			$\#$ `Dementia after Stroke' &  `Time after Stroke' & Point & SE \\ 
			$\leftindex_U{n}^c_{S^1D\bullet} (\tau)$ & $\int_0^{\tau} \leftindex_U{y}_{S^1 \bullet}(t) dt$   & $\hat{\lambda}_{S^1D}$ & $\sqrt{\mathcal{J}_{\tau}^{-1}(\hat{\lambda}_{S^1 D})}$ \\	\hline	
			\noalign{\smallskip}
			6275 & 85,645  & 0.072 & 0.00092 \\  \hline
			$\#$ `Dementia without Stroke'   &  `Healthy Times' & Point & SE \\
			$\leftindex_U{n}^c_{HD\bullet} (\tau)$  & $\int_0^{\tau} \leftindex_U{y}_{H \bullet}(t) dt$
			& $\hat{\lambda}_{HD}$ & $\sqrt{\mathcal{J}_{\tau}^{-1}(\hat{\lambda}_{HD})}$\\ \hline
			27,737  & 1,638,651 & 0.017 & 0.000102 \\ \hline   
			\noalign{\smallskip}       
	\end{tabular}}
\end{table}
Point estimates \eqref{ehaest} and standard errors (roots of \eqref{ehastand}) are given in Table \ref{stathomo}. Note that the 1,724,296 person-years at dementia risk of Table \ref{desstat} split into 85,645 after at stroke and 1,638,651 without a stroke. 
Even though 6275 dementia cases after a stroke do not appear to be very large, compared to the overall 25 thousand stroke cases, we find that a stroke increases the intensity of suffering from dementia from $\hat{\lambda}_{HD}$  $\approx$ 0.02 to $\hat{\lambda}_{S^1D}$ $\approx$ 0.07. Due to the asymptotic independence of both estimators, by Theorem \ref{normmodelb}, it is $Var(\hat{\lambda}_{S^1D}-\hat{\lambda}_{HD}) =  Var(\hat{\lambda}_{S^1D}) + Var(\hat{\lambda}_{HD})$ and (see Table \ref{stathomo}) estimated to be 0.00093$^2$ + 0.000102$^2$ = 8.8 $\times$ 10$^{-7}$. Hence, approximately $\hat{\lambda}_{S^1D}-\hat{\lambda}_{HD} \sim N(\lambda_{S^1D0}-\lambda_{HD0}, 0.000936^2)$, so that an approximate 95\%-confidence interval of the intensity difference is [0.455 $\pm$ 0.00183] = [0.050,0.056]. As the aim of the study is to answer the question whether having had a stroke has an effect on dementia onset, the corresponding Wald-test rejects, at level 5\%, because the confidence interval does not overlap with zero. This is equivalent to the statement that the absolute of the standardised intensity difference, as test statistics, exceeds the 97.5\% quantile of the normal distribution. The generalisation to the age-inhomogeneous model in Section \ref{dsm2}, will extend that, equivalently, the squared of that test statistics exceeds the 95\% quantile of the $\chi^2_1$ distribution.

To compare with, \cite{desmond2002} reveal an increased relative risk (RR) for dementia of 3.8 after a stroke, adjusted for several demographic factors and cognitive status. Note that the similarity of incidences and intensities, are argued in Section \ref{sec232}, enables to compare an RR, as ratio of incidences, directly with a ratio of intensities. Our unadjusted intensity ratio is 0.07/0.02=3.5. Adjustment for age will follow in Section \ref{dsm2}, and further for multi-morbidity in Section \ref{dsm3}.
Within the Framingham Study \citep{ivan2004}, the adjusted RR of dementia with respect to stroke is estimated to be 2.0. The Rotterdam Study \citep{reitz2008} also indicates that a stroke doubles the risk of dementia (hazard ratio:  HR=2.1). A systematic review and meta-analysis reveals a pooled HR of between 1.7 and 2.2 \citep{kuzma2018}. Another result, but without multi-states, is that of \cite{savva2010}, who report a hazard ratio of 2. Based on South Korean HCD, and also using multi-state methods, \cite{kimlee2018} find a 2.4-fold risk of subsequent dementia after a stroke. Our intensity ratio of 3.5 exceeds the more recent studies presumably because those adjust for covariates. We now as first covariate, we now adjust for age using age-inhomogeneous, namely piecewise constant, intensities. We will see that the effect of a stroke on dementia onset becomes markedly smaller because of Simpson's paradox: Simultaneously, intensities increase with age and a stroke is more likely at higher ages.

\section{\sc Adjustments for age and other diseases} \label{sec3}

In Section \ref{dsm1}, the probability of suffering either event, stroke or dementia onset, has been equal for all ages and independent of any other factor. Morbidity intensities vary with age and in order to compare our results for Germany later in Section \ref{sec33} internationally, we derive a model in Section \ref{dsm2} that adjusts for age inhomogeneity. 
Also, a risk-increasing effect of stroke on the dementia hazard might not be causal in the following sense. Assume that one group has a vascular predisposition and that a stroke (mainly) indicates the membership to that group. The information about the predisposition could have been achieved earlier and a stroke should not trigger additional medical effort with regard to dementia prevention. We aim in Section \ref{dsm3} at stratification according to vascular predisposition.  

\subsection{\sc Age-inhomogeneous intensities} \label{dsm2}

We define \citep[as in][]{testing-ho:2006,A-likeliho:2009}, for a partition $0=t_0, \ldots, t_b=\tau$, $\mathbf{X}$ as a Markov process with piece-wise constant intensities 
\begin{equation} \label{modpieces}
	\lambda_{hj}(t) := \sum_{l = 1}^b \mathds{1}_{[t_{l-1},t_l)}(t) \lambda_{hjl}.
\end{equation}
We do give neither the self-contained analysis of the lifetime state model of Appendix \ref{lsm}, nor the still complete analysis of the the age-homogeneous disease state model in Section \ref{dsm1}. We restrict the display to the statement of the conditional likelihood and derive of the estimator. The asymptotic arguments are developed to the extend that the standard errors can be derived. 

\subsubsection{Point estimate and standard error}\label{anaageinhom}

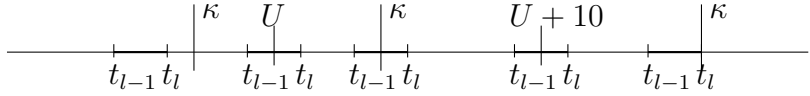
\begin{figure}[h]
		\begin{picture}(200, 40)
			\put(30.0, 20.0){\thinlines\line(1, 0){300.0}} 
			\put(130.0, 15.0){\thinlines\line(0, 1){15.0}} 
			\put(230.0, 15.0){\thinlines\line(0, 1){15.0}} 
			\put(125.0, 30.0){$U$}
			\put(218.0, 30.0){$U+10$}
			
			\put(70.0, 18.0){\thinlines\line(0, 1){5.0}}
			\put(70.0, 20.0){\thicklines\line(1, 0){20.0}} 
			\put(90.0, 18.0){\thinlines\line(0, 1){5.0}} 
			\put(68.0, 8.0){$t_{l-1}$}
			\put(88.0, 8.0){$t_l$}
			
			\put(120.0, 18.0){\thinlines\line(0, 1){5.0}}
			\put(120.0, 20.0){\thicklines\line(1, 0){20.0}} 
			\put(140.0, 18.0){\thinlines\line(0, 1){5.0}} 
			\put(118.0, 8.0){$t_{l-1}$}
			\put(138.0, 8.0){$t_l$}
			
			\put(160.0, 18.0){\thinlines\line(0, 1){5.0}}
			\put(160.0, 20.0){\thicklines\line(1, 0){20.0}} 
			\put(180.0, 18.0){\thinlines\line(0, 1){5.0}} 
			\put(158.0, 8.0){$t_{l-1}$}
			\put(178.0, 8.0){$t_l$}
			
			\put(220.0, 18.0){\thinlines\line(0, 1){5.0}}
			\put(220.0, 20.0){\thicklines\line(1, 0){20.0}} 
			\put(240.0, 18.0){\thinlines\line(0, 1){5.0}} 
			\put(218.0, 8.0){$t_{l-1}$}
			\put(238.0, 8.0){$t_l$}
			
			\put(270.0, 18.0){\thinlines\line(0, 1){5.0}}
			\put(270.0, 20.0){\thicklines\line(1, 0){20.0}} 
			\put(290.0, 18.0){\thinlines\line(0, 1){5.0}} 
			\put(268.0, 8.0){$t_{l-1}$}
			\put(288.0, 8.0){$t_l$}
			
			\put(170.0, 13.0){\thinlines\line(0, 1){25.0}} 
			\put(173.0, 33.0){$\kappa$}
			\put(100.0, 13.0){\thinlines\line(0, 1){25.0}} 
			\put(103.0, 33.0){$\kappa$} 
			\put(290.0, 13.0){\thinlines\line(0, 1){25.0}} 
			\put(293.0, 33.0){$\kappa$}  
			
		\end{picture}\caption{Possible intersections of $[t_l,t_{l-1})$ and $[U,U+10)$, possible situation for $\kappa$ in Section \ref{12345}} \label{Zeitstrahl}
\end{figure}

The same two counting processes $N_{hj}(t)$ and $Y_h(t)$ of Section \ref{modeldsm}, reformulate a history.
When stacking $\lambda_{hj}(t)$ to $\boldsymbol{\lambda}(t)$ in the same way as $N_{hj}$ to $\mathbf{N}_X$, $\mathbf{N}_X$ has a compensator - with respect to $\mathcal{N}_t$ - with intensity $\boldsymbol{\alpha}(t):=(Y_h(t) \lambda_{hj}(t), hj \in \mathcal{I})'$.
The compensator is with respect to the probability measure $\tilde{P}_{\boldsymbol{\lambda}}$, where $\boldsymbol{\lambda}:= (\lambda_{HS^11}, \ldots, \lambda_{Ddb})'$ collects the $6b$ parameters. Hence with little change, compared to \eqref{likeraneha}, the conditional likelihood contribution is
\begin{equation} \label{10}
	\begin{split}
		\ln \leftindex_U{L}^c(\mathbf{X},U|\boldsymbol{\lambda})  = & \int_U^{U+10} \sum_{hj \in \mathcal{I}} (\ln \leftindex_U{Y}_h(t) + \ln \lambda_{hj}(t)) d \leftindex_U{N}^c_{hj} (t) \\
		& - \sum_{hj \in \mathcal{I}} \int_U^{U+10} \leftindex_U{Y}_h(t) \lambda_{hj}(t) dt.
	\end{split} 
\end{equation}
Note that there are five possibilities for the intersection of $[t_l,t_{l-1})$ with $[U,U+10)$ (see Figure \ref{Zeitstrahl}), so that for the $N$-term (with $a \lor b := \max(a,b)$) for $[t_l,t_{l-1}) \cap [U,U+10) \neq \emptyset$ (0 elsewise)
\begin{multline*}
	\int_U^{U+10} \frac{\partial}{\partial \lambda_{hjl}} \ln \left( \sum_{\tilde{l}=1}^b  \mathds{1}_{[t_{\tilde{l}-1},t_{\tilde{l}})} (t) \lambda_{hj\tilde{l}} \right) d \leftindex_U{N}^c_{hj} (t) \\
	=	\int_U^{U+10} \frac{\mathds{1}_{[t_{l-1},t_l)} (t)}{\sum_{\tilde{l}=1}^b  \mathds{1}_{[t_{\tilde{l}-1},t_{\tilde{l}})} (t) \lambda_{hj\tilde{l}}} d \leftindex_U{N}^c_{hj} (t) \\
	= \int_{U \lor t_{l-1}}^{U+10 \land t_l} \frac{1}{\lambda_{hjl}} d \leftindex_U{N}^c_{hj} (t) 
	=	\frac{1}{\lambda_{hjl}} [\leftindex_U{N}^c_{hj} (U+10 \land t_l) - \leftindex_U{N}^c_{hj} (U \lor t_{l-1}) ]
\end{multline*}
and for the $Y$-term
\begin{multline*}
	-	\frac{\partial}{\partial \lambda_{hjl}}   \sum_{hj \in \mathcal{I}} \int_U^{U+10} \leftindex_U{Y}_h(t) \left( \sum_{\tilde{l}=1}^b  \mathds{1}_{[t_{\tilde{l}-1},t_{\tilde{l}})} (t) \lambda_{hj\tilde{l}} \right) dt  \\	= - \int_U^{U+10} \leftindex_U{Y}_h(t) \mathds{1}_{[t_{l-1},t_l)} (t) dt 	= - \int_{U \lor t_{l-1}}^{U+10 \land t_l} \leftindex_U{Y}_h(t) dt. 
\end{multline*}
Again by \eqref{likeraneha} with $\boldsymbol{\lambda}$, and comparable to \eqref{loglhomo}, it is $\ln \leftindex_U{L}^c(data;\boldsymbol{\lambda})$ the sum of the contributions \eqref{10}, so that  $ (\partial/\partial \lambda_{hjl}) \ln \leftindex_U{L}^c(data|\boldsymbol{\lambda})  =  A_{hj,l}/\lambda_{hjl} - B_{h,l}$, by interchanging differentiation and summation, with transitions $A_{hj,l}$ and time-at-risk $B_{h,l}$, per age-group, namely: 
\begin{eqnarray*}
	A_{hj,l} & := & \sum_{\stackrel{i=1}{[t_l,t_{l-1}) \cap [U_i,U_i+10) \neq \emptyset}}^n\left[\leftindex_U{N}^c_{hji} (U_i+10 \land t_l) - \leftindex_U{N}^c_{hji} (U_i \lor t_{l-1})\right] \\ 
	B_{h,l} & := & \sum_{\stackrel{i=1}{[t_l,t_{l-1})\cap [U_i,U_i+10) \neq \emptyset}}^n  \int_{U_i \lor t_{l-1}}^{U_i+10 \land t_l} \leftindex_U{Y}_{hi}(t) dt
\end{eqnarray*}  
Similar to \eqref{ehaest}, for the time interval $[t_l,t_{l-1}]$ it is:
\begin{equation} \label{ehaestai}
	\hat{\lambda}_{hjl} = A_{hj,l}/B_{h,l}
\end{equation} 
For the multi-state Markov model with right-censoring (but without left-trun\-cation), proof of the asymptotic normality (assuming consistency) for the piece-wise cons\-tant-intensity model \eqref{modpieces} is found in \cite{A-likeliho:2009}. A simplified proof for consistency is found in \cite{weismoll2011}. It is to be expected that the proofs easily generalise the case of left-truncation, because, similarly, a Doob-Meyer decomposition of the counting process into compensator and martingale is applied and enables the martingale limit theorem.  Hence, in order to derive confidence intervals only, agian, the Hessian of the logarithmic conditional likelihood is needed. It is a diagonal matrix with diagonal elements $- \mathcal{J}_{\tau}(\boldsymbol{\lambda}_0)_{hjl,hjl} = \frac{\partial^2}{\partial \lambda_{hjl}^2} \ln \leftindex_U{L}^c(data|\boldsymbol{\lambda})|_{\boldsymbol{\lambda}=\boldsymbol{\lambda}_0} 
=  - A_{hj,l}/\lambda_{hjl^0}^2$, and hence similar to \eqref{ehastand},
\begin{equation} \label{ehastand2}
	\widehat{Var(\hat{\lambda}_{hjl})} =  A_{hj,l}/B_{h,l}^2. 
\end{equation}


\subsubsection{Result for AOK HCD} \label{sec33}

\begin{figure}[t]
	\centering
	\includegraphics[scale=0.85]{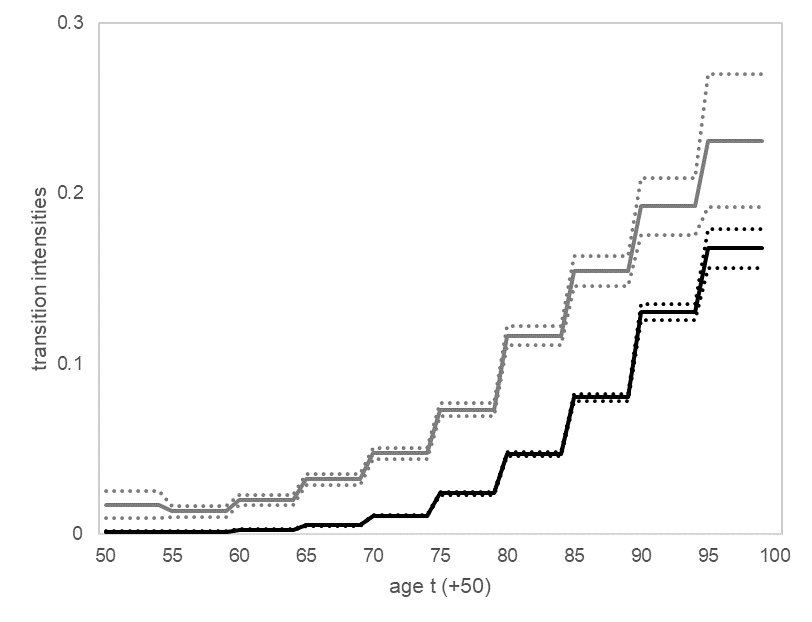}
	\caption{Age-inhomogeneous intensity from stroke ($S^1$) to dementia onset ($D$) 	$\hat{\lambda}_{S^1D}(t)$ (top, grey) and healthy ($H$) to dementia onset ($D$) $\hat{\lambda}_{HD}(t)$ (bottom, black) on 5-year intervals: Point estimate (\eqref{ehaestai}, solid line) and 95\% confidence interval (\eqref{ehastand2}, dashed line)}
	\label{logLQ_a1}
\end{figure}

Population and data, including the number of observation $n$, all remain the same as in age-homogeneous model of Section \ref{dsm1}. Section \ref{res2} found an effect of stroke on dementia onset that exceeds by far findings in contemporary epidemiology. An age-inhomogeneous dementia intensity has already been confirm for the AOK HCD in \cite{weisetal2020} and we now apply the piece-wise constant intensities \eqref{modpieces}. Table \ref{statinhomo} and Figure \ref{logLQ_a1} exhibit point estimates, standard errors and confidence intervals due to \eqref{ehaestai}, \eqref{ehastand2} and the generalisation of Theorem \ref{normmodelb} with age intervals covering five years, i.e. with {\em b} = 12 pieces (see Table \ref{statinhomo}, column (1)). 
\begin{sidewaystable} 
	\centering
	\caption{Statistics (columns (2),(3),(5),(6)), point estimates with standard errors (SE) (Formulae: \eqref{ehaestai} and  root of \eqref{ehastand2}) (columns (4),(7)) and intensity ratio (column (8)) for 5-year age-classes (Oldest class [110-114) has no stroke or dementia events)}
	\label{statinhomo} \smallskip
	{\footnotesize
		\begin{tabular}{cccccccc}
			\hline \hline
			\noalign{\smallskip}
			(1) & (2) & (3) & (4) & (5) & (6) & (7) & (8) \\
			Age interval & $\#$ `With dementia   &  `Time after stroke & Point & $\#$ `With dementia    &  `Healthy times & Point & Intensity \\ 
			$[t_{l-1}+ 50,t_l+50)$ & after stroke' & in age interval'  & $\hat{\lambda}_{S^1Dl}^{\pm SE}$ & without stroke'  & in age interval'	& $\hat{\lambda}_{HDl}^{\pm SE}$ & ratio (4)/(7) \\	\hline			
			\noalign{\smallskip}			
			$[$50,55)	&	19	&	1103	&	0.0172$^{\pm 0.0040}$ &	128	&	95939	&	0.0013$^{\pm 0.0001}$ & 12.9  \\
			$[$55,60)	&	65	&	4874	&	0.0133$^{\pm 0.0017}$ 	&	322	&	227939	&	0.0014$^{\pm 0.0001}$ & 9.4	  \\
			$[$60,65)	&	158	&	7970	&	0.0198$^{\pm 0.0016}$ 	&	638	&	250821	&	0.0025$^{\pm 0.0001}$ & 7.8  \\
			$[$65,70)	&	382	&	11911	&	0.0321$^{\pm 0.0016}$ 	&	1469	&	287148	&	0.0051$^{\pm 0.0001}$  & 6.3 \\
			$[$70,75)	&	828	&	17515	&	0.0473$^{\pm 0.0016}$  	&	3222	&	298790	&	0.0108$^{\pm 0.0002}$ & 4.4 \\
			$[$75,80)	&	1306	&	17876	&	0.0731$^{\pm 0.0020}$  	&	5497	&	230546	&	0.0238$^{\pm 0.0003}$  & 3.1	 \\
			$[$80,85)	&	1675	&	14381	&	\fbox{0.1165$^{\pm 0.0028}$} 	&	7082	&	150258	&	\fbox{0.0471$^{\pm 0.0006}$} & 2.5	  \\
			$[$85,90)	&	1192	&	7730	&	0.1542$^{\pm 0.0045}$ 	&	5575	&	69521	&	0.0802$^{\pm 0.0011}$   & 1.9 \\
			$[$90,95)	&	506	&	2629	&	0.1925$^{\pm 0.0086}$ 	&	2900	&	22216	&	0.1305$^{\pm 0.0024}$  &	1.5  \\
			$[$95,100)	&	136	&	589	&	0.2309$^{\pm 0.0198}$ 	&	826	&	4927	&	0.1677$^{\pm 0.0058}$ & 1.4	  \\
			$[$100,105)	&	8	&	66	&	0.1208$^{\pm 0.0427}$ 	&	86	&	525	&	0.1640$^{\pm 0.0177}$ & 0.7	  \\
			$[$105,110)	&	0	&	0{.}5	&	0$^{\pm 0}$ 	&	2	&	22	&	0.0918$^{\pm 0.0649}$ & NA  \\	\hline
			$\sum$ 		&	6275 \text{events}	&	\multicolumn{2}{l}{86,645 \; \text{person-years at risk}}		&	27,737 \text{events}	&	\multicolumn{2}{l}{1{,}638{,}651 \; \text{person-years at risk}} 	&   \\	\hline \hline
			\noalign{\smallskip}       
	\end{tabular}}
\end{sidewaystable}
For instance, in the age-group with the most dementia events, namely from 80 to 85 years, the dementia intensity after stroke of $\hat{\lambda}_{S^1D7}$ = 0.117 exceeds that without stroke of $\hat{\lambda}_{HD7}$ = 0.047 (see framed numbers in Table \ref{statinhomo}). The ratio of 2.5 is now two thirds of the ratio 0.07/0.02 = 3.5 of Table \ref{stathomo}, and more in line with the recent literature, e.g. for Korea, as reported by \cite{kimlee2018} (of 2.4). The reason is Simpson's paradox; the age-homogeneous $\hat{\lambda}_{S^1 D}$ = 0.07 of Section \ref{res2} is implicitly an average over a later part of the time span $[0,\tau]$ where dementia onset is anyway more likely. In more detail, a stroke generally occurs at higher ages, so that the denominator in the stroke-specific point estimator \eqref{ehaest}, starts accumulating `time at risk' at a high age. The higher dementia intensity at the ages then results in many events in the numerator of point estimator \eqref{ehaest}, not attributable to the stroke event. This defect is resolved by the age-specific ratios in \eqref{ehaestai}. And the defect does not level out when calculating the relative risk, because the defect does not affect the healthy persons' intensity $\hat{\lambda}_{HD}$. The formulated aim of the study is to answer the question whether stroke has a effect on dementia. Following up on the arguments in Section \ref{res2}, consider the squared test statistics for each of the $b$ = 12 time intervals, and add those. Thus sum for the 12 differences is distributed as $\chi^2_{12}$ due to the independence between estimation differences, which must hold in analogy to Theorem \ref{normmodelb} also for age-inhomogeneity as the proof of \citet[][Theorem 1]{A-likeliho:2009} suggests. The 95\% quantile of the $\chi^2_{12}$-distribution is 21.026 and the test statistic (using \eqref{ehastand2})
\[
\sum_{l=1}^b \left( \frac{\hat{\lambda}_{S^1Dl}-\hat{\lambda}_{HDl}}{ \sqrt{\widehat{Var(\hat{\lambda}_{S^1Dl})} +  \widehat{Var(\hat{\lambda}_{HDl})}}} \right)^2=0.289,
\]
so that the test is significant.  

In order to explore the role of age further, we may notice a decreasing stroke-effect in age, measured in ratios. In the age group of the 55 to 60 years old, the intensity ratio is 9.4 (see last column in Table \ref{statinhomo}). The higher the age is, the smaller is the intensity ratio. This coincides with the Framingham Study \citep{ivan2004} where the adjusted RR was higher for those younger than 80 (RR=2.6) compared to those aged 80 or older (RR=1.6). The $b$ = 12 age-specific Wald-type tests for pairwise differences (suppressed here) show that there is no significant difference in the risk of dementia between persons with and without stroke for the highest age groups (90 years and older). Similarly, the systematic review by \cite{savva2010} also does not find an excess risk of dementia after stroke in those at ages 85 years or older.

\subsection{\sc Stratification by multi-morbidity} \label{dsm3}

We now stratify according to vascular predisposition. Six diseases (other than stroke) are considered as potential vascular risk factors mentioned in \cite{mangia2012}. Hypertension (ICD-10: I10-I15) is the most frequent condition with a prevalence of approximately 90\% in the data. Hence, itself does not stratify sufficiently, so that we use the ``two out of six'' rule including the five other factors ``type 2 diabetes'' (ICD-10: E11-E14), ``ischemic heart diseases'' (ICD-10: I20-I25), ``atrial fibrillation'' (ICD-10: I48), ``hypercholesterolemia'' (ICD-10: E78.0) and ``obesity'' (ICD-10: E66). 

Stratification according to multi-morbidity at the time origin, i.e. at the age of 50, would impose a random dichotomous classifier $Z$, but is not observed as some people are older in 2004. Moreover, multi-morbidity is age dependent, as acquisition of the second vascular disease could take place at any age after age 50. A time-dependent covariate $Z(t)=\mathds{1}_{\{[\text{age at the second disease onset},\tau)\}}(t)$ is necessary:
For a dichotomous covariate, the additive model $\lambda^{wo}_{hjl}+ \tilde{\beta}_{hjl}Z(t)$, as in \citet[][Formula 5]{kremweis2013}, or the multiplicative $\lambda^{wo}_{hjl}e^{\beta_{hjl}Z(t)}$ \cite[][Formular 7.6.2]{And} are equal and we may write the model as piece-wise constant.  Theory for an additive model and a fixed $z$ is derived in \cite{kremweis2013}, for a lifetime state model with left-censoring. For the multiplicative model and right-censoring, \citet[][Theorem 2]{Bor} derives the asymptotic distribution of the estimator.  The full theory for left-truncation will not be be reported here, only the point estimator and standard error shall be given. Observable data require a random $Z$, as usual, and we assume that the distribution of $Z$ does ot depend on $\boldsymbol{\lambda}$ and condition again (after conditioning on $U$ and $X_0$) on $Z$. For the ease of notation, define the age of multi-morbidity onset as $\kappa:=\min \{t: z(t)=1\}$. We refrain from developing an age-homogeneous model and directly follow up on Section \ref{dsm2} model age-inhomogeneously.

\subsubsection{Point estimate and standard error} \label{12345}
For each person only one additional split on the constant intensities pieces is necessary. For (the random version of) $\kappa$ before $U$ or after $U$ + 10, no further distinction is necessary (see Figure \ref{Zeitstrahl}). The idea is that a person now contributes to the estimation of a set of parameters without multi-morbidity, $\lambda^{wo}_{hjl}$,  , i.e. to the transition counts and the at-risk-times, until that $\kappa$. After the split, a set of parameters with multi-morbidity, $\lambda^w_{hjl}$, 
is estimated. All parameters are collected in $\boldsymbol{\lambda}$. In detail, conditional on $Z=z$ we define
\begin{equation} \label{zmodelmm}
	\lambda_{hj}(t,z(t))
	:=  \sum_{l = 1}^b \mathds{1}_{[t_{l-1},t_l)}(t) \left(\mathds{1}_{[0,\kappa)}(t) \lambda^{wo}_{hjl}+ \mathds{1}_{[\kappa, \infty)}(t)) \lambda^{w}_{hjl} \right).
\end{equation}  
Obviously, generalising \eqref{likeraneha}, as in \eqref{10}, the logarithmic conditional likelihood contribution is
\begin{equation} \label{ztloglik}  
	\begin{split}
		\ln \leftindex_U{L}^c(\mathbf{X},U,\mathbf{Z}|\boldsymbol{\lambda})  = & \int_U^{U+10} \sum_{hj \in \mathcal{I}} (\ln \leftindex_U{Y}_h(t) + \ln \lambda_{hj}(t,z(t))) d \leftindex_U{N}^c_{hj} (t) \\
		& - \sum_{hj \in \mathcal{I}} \int_U^{U+10} \leftindex_U{Y}^c_h(t) \lambda_{hj}(t,z(t)) dt
	\end{split} 
\end{equation}
Now, in extension to Section \ref{anaageinhom}, for $\kappa \in [U,U+10)$, the derivative of the $N$-term of $ \frac{\partial}{\partial \lambda^{wo}_{hjl}} \ln \leftindex_U{L}^c(\mathbf{X},U,\mathbf{Z}|\boldsymbol{\lambda})$ is
{\footnotesize
	\begin{multline*}
		\int_U^{U+10} \frac{\partial}{\partial \lambda^{wo}_{hjl}} \ln \left(  \sum_{\tilde{l} = 1}^b \mathds{1}_{[t_{\tilde{l}-1},t_{\tilde{l}})}(t) \left(\mathds{1}_{[0,\kappa)}(t) \lambda^{wo}_{hj\tilde{l}} + \mathds{1}_{[\kappa, \infty)}(t) \lambda^{w}_{hj\tilde{l}} \right) \right) d \leftindex_U{N}^c_{hj} (t) \\
		=	\int_U^{U+10} \frac{\mathds{1}_{[t_{l-1},t_l) \cap [0,\kappa)} (t)  }{ \sum_{\tilde{l} = 1}^b \mathds{1}_{[t_{\tilde{l}-1},t_{\tilde{l}})}(t) \left(\mathds{1}_{[0,\kappa)}(t) \lambda^{wo}_{hj\tilde{l}}+ \mathds{1}_{[\kappa, \infty)}(t) \lambda^{w}_{hj\tilde{l}} \right)} d \leftindex_U{N}^c_{hj} (t) \\
		= \left\{ \mbox{\begin{tabular}{ll} $0$ & \text{if} \; $\kappa \le t_{l-1}$  \\
				$\frac{1}{\lambda^{wo}_{hjl}} \int_U^{U+10} \mathds{1}_{[t_{l-1},t_l)} (t) d \leftindex_U{N}^c_{hj} (t)= \frac{1}{\lambda^{wo}_{hjl}}(\leftindex_U{N}^c_{hj} (t_l) - \leftindex_U{N}^c_{hj} (t_{l-1}))$ & \text{if} \; $\kappa \ge  t_l$ \\				 
				$\frac{1}{\lambda^{wo}_{hjl}} \int_U^{U+10}  \mathds{1}_{[t_{l-1},\kappa)} (t) d \leftindex_U{N}^c_{hj} (t)=\frac{1}{\lambda^{wo}_{hjl}}(\leftindex_U{N}^c_{hj} (\kappa) - \leftindex_U{N}^c_{hj} (t_{l-1}))$ & 	\quad \text{if} \; $\kappa \in [t_{l-1}, t_l]$ 
		\end{tabular}} \right. 
	\end{multline*}
}
and for $ \frac{\partial}{\partial \lambda^w_{hjl}} \ln \leftindex_U{L}^c(\mathbf{X},U,\mathbf{Z}|\boldsymbol{\lambda})$ for the $N$-term it is:
{\footnotesize
	\begin{multline*}
		\int_U^{U+10} \frac{\partial}{\partial \lambda^w_{hjl}} \ln \left(  \sum_{\tilde{l} = 1}^b \mathds{1}_{[t_{\tilde{l}-1},t_{\tilde{l}})}(t) \left(\mathds{1}_{[0,\kappa)}(t) \lambda^{wo}_{hj\tilde{l}} + \mathds{1}_{[\kappa, \infty)}(t) \lambda^{w}_{hj\tilde{l}} \right) \right) d \leftindex_U{N}^c_{hj} (t) \\
		=	\int_U^{U+10} \frac{\mathds{1}_{[t_{l-1},t_l) \cap [\kappa,\infty)} (t)  }{ \sum_{\tilde{l} = 1}^b \mathds{1}_{[t_{\tilde{l}-1},t_{\tilde{l}})}(t) \left(\mathds{1}_{[0,\kappa)}(t) \lambda^{wo}_{hj\tilde{l}}+ \mathds{1}_{[\kappa, \infty)}(t) \lambda^{w}_{hj\tilde{l}} \right)} d \leftindex_U{N}^c_{hj} (t) \\
		= \left\{ \mbox{\begin{tabular}{ll} 	$ \frac{1}{\lambda^w_{hjl}}\int_U^{U+10}  \mathds{1}_{[t_{l-1},t_l)} (t) d \leftindex_U{N}^c_{hj} (t)=\frac{1}{\lambda^w_{hjl}}(\leftindex_U{N}^c_{hj} (t_l) - \leftindex_U{N}^c_{hj} (t_{l-1}))$ & \text{if} \; $\kappa \le t_{l-1}$  \\
				$0$ & \text{if} \; $\kappa \ge  t_l$ \\				 
				$\frac{1}{\lambda^w_{hjl}} \int_U^{U+10} \mathds{1}_{[\kappa,t_l)} (t) d \leftindex_U{N}^c_{hj} (t)=\frac{1}{\lambda^w_{hjl}}(\leftindex_U{N}^c_{hj} (t_l) - \leftindex_U{N}^c_{hj} (\kappa))$ & 	 \text{if} \; $\kappa \in [t_{l-1}, t_l]$ 
		\end{tabular}} \right. 
	\end{multline*} 
}
For $\kappa \ge U+10$, the $N$-term in
$ \frac{\partial}{\partial \lambda^{wo}_{hjl}} \ln \leftindex_U{L}^c(\mathbf{X},U,\mathbf{Z}|\boldsymbol{\lambda})$ is $ (\lambda^{wo}_{hjl})^{-1}(\leftindex_U{N}^c_{hj} (t_l) - \leftindex_U{N}^c_{hj} (t_{l-1}))$ and the derivative with respect to $\lambda^w_{hjl}$ is zero.   For $\kappa < U$, the $N$-term with respect to $\lambda^{wo}_{hjl}$ zero and in 
$ \frac{\partial}{\partial \lambda^w_{hjl}} \ln \leftindex_U{L}^c(\mathbf{X},U,\mathbf{Z}|\boldsymbol{\lambda})$ it is $ (\lambda^w_{hjl})^{-1}(\leftindex_U{N}^c_{hj} (t_l) - \leftindex_U{N}^c_{hj} (t_{l-1}))$.

Now, the derivative of the $Y$-term of $ \frac{\partial}{\partial \lambda^{wo}_{hjl}} \ln \leftindex_U{L}^c(\mathbf{X},U,\mathbf{Z}|\boldsymbol{\lambda})$ is, in extension to Section \ref{anaageinhom}:
\begin{multline*}
	\frac{\partial}{\partial \lambda^{wo}_{hjl}}   \sum_{hj \in \mathcal{I}} \int_U^{U+10} \leftindex_U{Y}_h(t) \left( \sum_{\tilde{l} = 1}^b \mathds{1}_{[t_{\tilde{l}-1},t_{\tilde{l}})}(t) \left(\mathds{1}_{[0,\kappa)}(t) \lambda^{wo}_{hj\tilde{l}}+ \mathds{1}_{[\kappa, \infty)}(t) \lambda^{w}_{hj\tilde{l}} \right) \right) dt  \\	=  \int_U^{U+10} \leftindex_U{Y}_h(t) \mathds{1}_{[t_{l-1},t_l) \cap [0,\kappa)} (t)  dt \\				
	= \left\{ \mbox{\begin{tabular}{ll} 	$0$ & $\kappa \in [U,U+10)\; \text{and} \; \kappa \le t_{l-1}$  \\ 
			$\int_{t_{l-1}}^{\kappa} \leftindex_U{Y}_h(t) dt$ & \text{if} \; $\kappa \in [U,U+10)\; \text{and} \; \kappa \in [t_{l-1}, t_l]$ \\
			$\int_{t_{l-1}}^{t_l} \leftindex_U{Y}_h(t) dt$ & \text{if} \; $\kappa \in [U,U+10)\; \text{and} \; \kappa \ge  t_l$ \\				 
			$0$ & \text{if} \; $\kappa <  U$ \\	
			$\int_{t_{l-1}}^{t_l} \leftindex_U{Y}_h(t) dt$ & \text{if} \; $\kappa \ge  U+10$ \\	
	\end{tabular}} \right. \\
\end{multline*}
The $Y$-term of $ \frac{\partial}{\partial \lambda^w_{hjl}} \ln \leftindex_U{L}^c(\mathbf{X},U,\mathbf{Z}|\boldsymbol{\lambda})$ is:
\begin{multline*}
	\frac{\partial}{\partial \lambda^w_{hjl}}   \sum_{hj \in \mathcal{I}} \int_U^{U+10} \leftindex_U{Y}_h(t) \left( \sum_{\tilde{l} = 1}^b \mathds{1}_{[t_{\tilde{l}-1},t_{\tilde{l}})}(t) \left(\mathds{1}_{[0,\kappa)}(t) \lambda^{wo}_{hj\tilde{l}}+ \mathds{1}_{[\kappa, \infty)}(t) \lambda^{w}_{hj\tilde{l}} \right) \right) dt  \\	=  \int_U^{U+10} \leftindex_U{Y}_h(t) \mathds{1}_{[t_{l-1},t_l) \cap [\kappa, \infty)} (t)  dt \\				
	= \left\{ \mbox{\begin{tabular}{ll}  $\int_{t_{l-1}}^{t_l} \leftindex_U{Y}_h(t) dt$	 & $\kappa \in [U,U+10)\; \text{and} \; \kappa \le t_{l-1}$  \\ 
			$\int_{\kappa}^{t_l} \leftindex_U{Y}_h(t) dt$ & \text{if} \; $\kappa \in [U,U+10)\; \text{and} \; \kappa \in [t_{l-1}, t_l]$ \\
			$0$ & \text{if} \; $\kappa \in [U,U+10)\; \text{and} \; \kappa \ge  t_l$ \\				 
			$\int_{t_{l-1}}^{t_l} \leftindex_U{Y}_h(t) dt$  & \text{if} \; $\kappa <  U$ \\	
			$0$ & \text{if} \; $\kappa \ge  U+10$ \\	
	\end{tabular}} \right. \\
\end{multline*}
These summarized, as in \eqref{loglhomo},  to
$\ln \leftindex_U{L}_{\tau}^c(data;\boldsymbol{\lambda})  =  \sum_{i=1}^n  \ln \leftindex_U{L}_{\tau}^c(\mathbf{X}^i,U_i,\mathbf{Z}_i|\boldsymbol{\lambda})$, and set to zero, results in
point estimators
\begin{equation} \label{eststrat}
	\hat{\lambda}^{wo}_{hjl}= A^{wo}_{hj,l}/B^{wo}_{h,l} \quad \text{and} \quad 	\hat{\lambda}^w_{hjl}= A^w_{hj,l}/B^w_{h,l} \quad \text{with}
\end{equation}
\begin{eqnarray*}
	A^{wo}_{hj,l} & := & \sum_{i=1}^n (\leftindex_U{N}^c_{hji} ((\kappa_i \land t_l) \lor t_{l-1}) - \leftindex_U{N}^c_{hji} (t_{l-1})),\\
	A^w_{hj,l} & := & \sum_{i=1}^n (\leftindex_U{N}^c_{hji} (t_l) - \leftindex_U{N}^c_{hji} ((t_{l-1} \lor \kappa_i) \land t_l)),\\
	B^{wo}_{h,l} & := & \sum_{i=1}^n  \int_{t_{l-1}}^{(t_l \land \kappa_i)\lor t_{l-1}} \leftindex_U{Y}_{hi}(t) dt
	\quad \text{and} \\		B^w_{h,l}  & := & \sum_{i=1}^n \int_{(t_{l-1} \lor \kappa_i) \land t_l}^{t_l} \leftindex_U{Y}_{hi}(t) dt .
\end{eqnarray*}
Their squared standard errors are, similar to \eqref{ehastand2}, $A^{wo}_{hj,l}/(B^{wo}_{h,l})^2$ and $A^w_{hj,l}/(B^w_{h,l})^2$.

\subsubsection{Result for AOK HCD}

Incorporating multi-morbidity, \eqref{zmodelmm}, two tables, similar to the unstratified Table \ref{statinhomo}, for the two groups with and without multi-morbidity are now given jointly in Table \ref{statinhomo2}. Comparing the columns 4 and 7, the dementia onset intensity is again larger when having had a stroke, as in the age-homogeneous model (of Section \ref{dsm1}) and in the age-inhomogeneous model (of Section \ref{dsm2}). Comparing the first and second rows, multi-morbidity does increase the dementia intensity, however much less than a stroke does.  

\begin{figure}[t]
	\centering
	\includegraphics[scale=0.337]{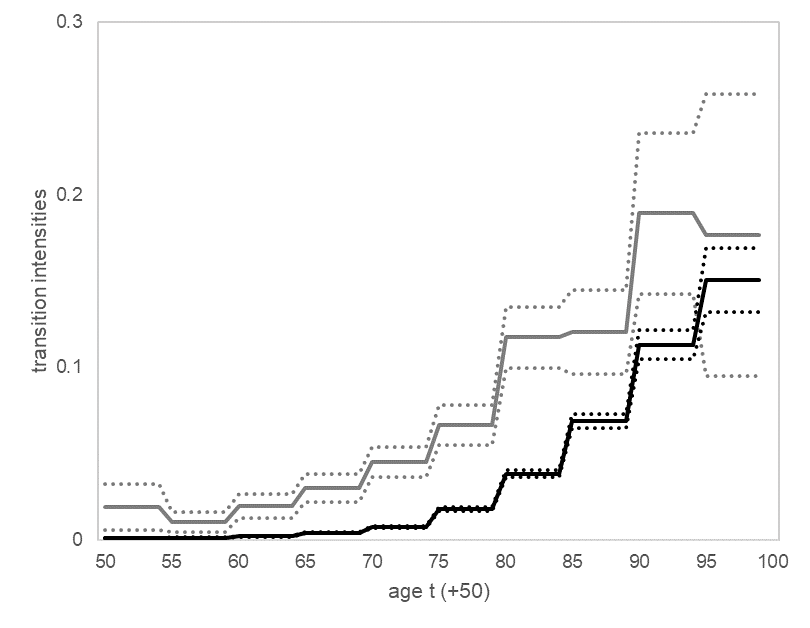}
	\includegraphics[scale=0.337]{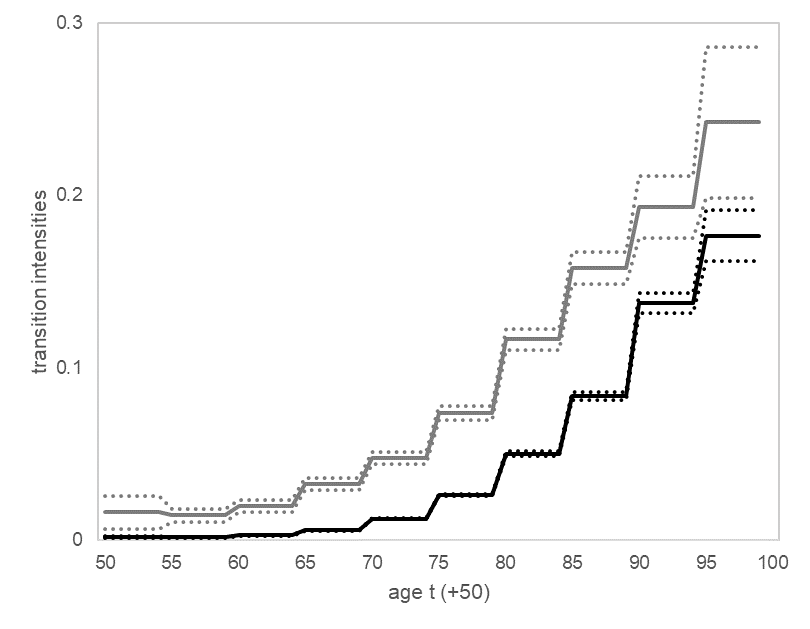}
	\includegraphics[scale=0.295]{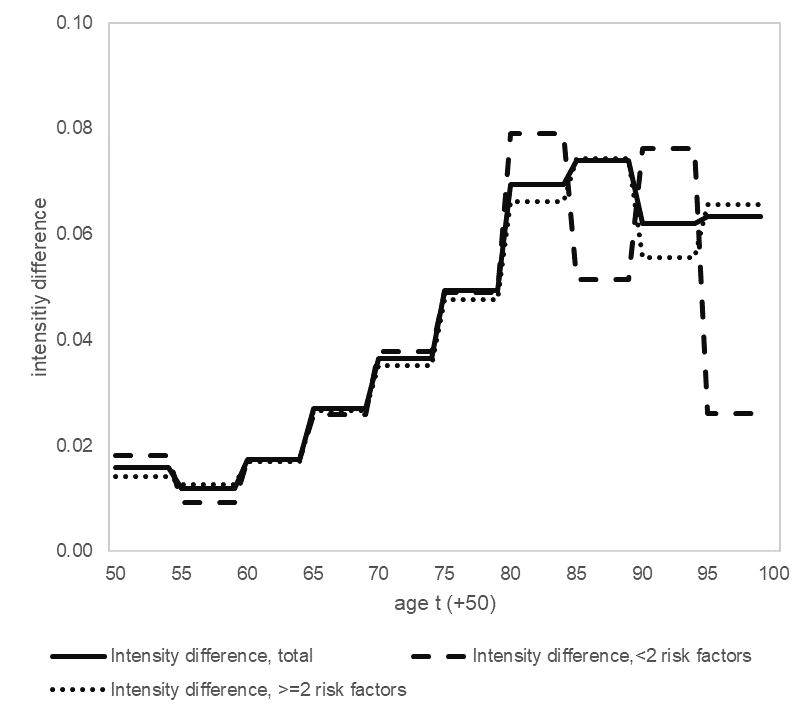}
	\caption{Estimates \eqref{eststrat} for intensity model \eqref{zmodelmm} as in Table \ref{statinhomo2} without multi-morbidity (left) from stroke ($S^1$) to dementia onset ($D$) 	$\lambda^{wo}_{S^1Dl}$ (top, grey) and healthy ($H$) to dementia onset ($D$) $\hat{\lambda}^{wo}_{HDl}$ (bottom, black), intensity with multi-morbidity $\lambda^w_{S^1Dl}$ and $\hat{\lambda}^w_{HDl}$ (middle); and differences thereof $\lambda^{wo}_{S^1Dl}-\lambda^{wo}_{HDl}$ (dashed), $\lambda^w_{S^1Dl}-\lambda^w_{HDl}$ (dotted) (right), combined with un-stratified from Table \ref{statinhomo} $\lambda_{S^1Dl}-\lambda_{HDl}$ (solid)}
	\label{fig6}
\end{figure}

\begin{sidewaystable} 
	\centering
	\caption{Statistics as in Table \ref{statinhomo} for point estimates \eqref{eststrat} and standard errors (SE), 1$^{st}$ row without multi-morbidity $\hat{\lambda}^{wo}_{hjl}$, 2$^{nd}$ row with multi-morbidity $\hat{\lambda}^w_{hjl}$ (result above age 100 are too small to be reported)}
	\label{statinhomo2} \smallskip
	{\footnotesize
		\begin{tabular}{cccccccc}
			\hline \hline
			\noalign{\smallskip}
			Age interval & $\#$ `With dementia   &  `Time after stroke & Point & $\#$ `With dementia    &  `Healthy times & Point & Intensity \\ 
			$[t_{l-1}+ 50,t_l+50)$ & after stroke' & in age interval'  & $\lambda_{S^1Dl}^{wo/w, \pm SE}$ & without stroke'  & in age interval'	& $\lambda_{HDl}^{wo/w, \pm SE}$ & difference (4)-(7) \\	\hline			
			\noalign{\smallskip}			
			$[$50,55)	&	8	&	418	&	0.019$^{\pm 0.0068}$ &	80	&	71669	&	0.0011$^{\pm 0.0001}$ & 0.0180  \\
			&	11	&	685	&	0.0160$^{\pm 0.0048}$ &	48	&	24270	&	0.0020$^{\pm 0.0003}$ & 0.0141  \\
			$[$55,60)	&	13	&	1254	&	0.0104$^{\pm 0.0029}$ 	&	167	&	135902	&	0.0012$^{\pm 0.0001}$ & 0.0091	  \\
			&	52	&	3621	&	0.0144$^{\pm 0.0020}$ 	&	155	&	92037	&	0.0017$^{\pm 0.0001}$ & 0.0127	  \\
			$[$60,65)	&	30	&	1531	&	0.0196$^{\pm 0.0036}$ 	&	279	&	124899	&	0.0022$^{\pm 0.0001}$ & 0.0174  \\
			&	128	&	6439	&	0.0199$^{\pm 0.0018}$ 	&	359	&	125922	&	0.0029$^{\pm 0.0002}$ & 0.0170  \\
			$[$65,70)	&	56	&	1861	&	0.0301$^{\pm 0.0040}$ 	&	528	&	124122	&	0.0043$^{\pm 0.0002}$  & 0.0258 \\
			&	326	&	10050	&	0.0324$^{\pm 0.0018}$ 	&	941	&	163025	&	0.0058$^{\pm 0.0002}$  & 0.0267 \\
			$[$70,75)	&	101	&	2232	&	0.0452$^{\pm 0.0045}$  	&	770	&	102328	&	0.0075$^{\pm 0.0003}$ & 0.0377 \\
			&	727	&	15283	&	0.0476$^{\pm 0.0018}$  	&	2452	&	196462	&	0.0125$^{\pm 0.0003}$ & 0.0351 \\
			$[$75,80)	&	129	&	1931	&	0.0668$^{\pm 0.0059}$  	&	1187	&	66273	&	0.0179$^{\pm 0.0005}$  & 0.0489	 \\
			&	1177	&	15945	&	0.0738$^{\pm 0.0022}$  	&	4300	&	164273	&	0.0262$^{\pm 0.0004}$  & 0.0476 \\
			$[$80,85)	&	167	&	1422	&	0.1174$^{\pm 0.0091}$ 	&	1474	&	38335	&	0.0385$^{\pm 0.0010}$ & 0.0790	  \\
			&	1508	&	12595	&	0.1164$^{\pm 0.0030}$ 	&	5608	&	111922	&	0.0501$^{\pm 0.0007}$ & 0.0663	  \\
			$[$85,90)	&	93	&	772	&	0.1205$^{\pm 0.0125}$ 	&	1137	&	16465	&	0.0691$^{\pm 0.0020}$   & 0.0514 \\
			&	1099	&	6959	&	0.1579$^{\pm 0.0048}$ 	&	4438	&	53056	&	0.0836$^{\pm 0.0013}$   & 0.0743 \\
			$[$90,95)	&	63	&	333	&	0.1894$^{\pm 0.0239}$ 	&	709	&	6367	&	0.1131$^{\pm 0.0042}$  &	0.0762  \\
			&	443	&	2296	&	0.1929$^{\pm 0.0092}$ 	&	2191	&	15948	&	0.1374$^{\pm 0.0029}$  &	0.0555  \\
			$[$95,100)	&	18	&	102	&	0.1768$^{\pm 0.0417}$ 	&	257	&	1705	&	0.1507$^{\pm 0.0094}$ & 0.0261	  \\
			&	118	&	487	&	0.2423$^{\pm 0.0223}$ 	&	569	&	3222 &	0.1766$^{\pm 0.0074}$ & 0.0657	  \\
			\hline \hline
			\noalign{\smallskip}       
	\end{tabular}}
\end{sidewaystable}
The graphical analysis of the estimates \eqref{eststrat} and confidence intervals (as in Figure \ref{logLQ_a1}) are displayed in Figure  \ref{fig6}. If multi-morbidity were a predominant predictive factor, a stroke would now not increase the dementia incidence. This is not the case as the middle panel shows.   
The two panels (left and middle) reveal little differences in dementia intensity of the stratification (apart from larger confidence intervals because group sizes are smaller than in  Figure \ref{logLQ_a1}). The differences $\lambda^{wo}_{S^1Dl}-\lambda^{wo}_{HDl}$ and $\lambda^w_{S^1Dl}-\lambda^w_{HDl}$ are equal, and also equal to the unstratified difference $\lambda_{S^1Dl}-\lambda_{HDl}$ (right panel) and hereby strongly suggest that stroke is a risk factor irrespective of multi-morbidity.   
It it tempting to construct a $\chi^2$-test for the global hypothesis that stroke is a significant risk factor, similar to that at the end of Section \ref{sec33}. However, it is not clear that the diagonal structure of the asymptotic variance-covariance matrix holds. Recall that in the linear regression, a regressor introduces dependence between the parameter estimators \cite[see e.g.][Sect. 21.2 or 23.3]{Bley2021}. 

\section{Conclusion}
\label{sec4}

Note first that left-truncation can be circumvented by matching the starts of population and observation period, i.e. hereby defining the population as those born after 1954, and hence turning 50 years old from 2004 onwards  \cite[as e.g. done in][]{testing-ho:2006}. However, not only will then the (many) events of stroke and dementia-onset for people born before 1954 be un-analysed, also will the population not be of current interest, because dementia is a disease of old-age. More critical is that the similarity of the stroke effect from Section \ref{dsm2} with that of the related study \cite{kimlee2018} for Korea is misleading because the later study takes more covariates into account. Effect sizes typically decrease as a function of the  number of covariates due to multicollinearity. However, integrating exogenous continuous covariates in our left-truncated event history analysis, other than the dichotomous covariate we considered,  is algorithmically cumbersome \cite[see e.g.][]{kimjame2011}. Also critical is that we assume three sorts of independence. First of all, we assume it {\em within} pairs $(T_i,U_i)$, even though this is likely to be untrue in our case study because $U$ is an affine transformation of the birth date. Typically demographers assume that younger cohorts tend to live longer. Theory about dependent truncation is currently developed in \cite{tenzer2021,UnaKeil2021,renxie2021}. For dependent truncation, the truncation time distribution becomes influential and different distributions are studied in \cite{weisdoer2022}. Second, we can assume independence {\em between} pairs $(T_i,U_i)$, due to our data being a sample. However, for an observational study of event histories, stochastic independence must be concluded from unforeseeable birth dates. Third, close to longitudinal independence, probably the most critical assumption of our modelling strategy seems to be the Markovian. This is especially the case, as for our application \cite{pend2009} and \cite{corraini2017} claim that the time elapsed since stroke is a risk factor for the intensity of dementia onset. Such duration-dependence especially violates the assumption of multiplicative intensities \eqref{multiintensity} and thus requires a different strategy \cite[see e.g.][]{weisschm2019}. Another critical point is that, after the first age axis time-since-birth and the second axis time-since-observation-start, a potential third axis is the cohort trend, found by \cite{weisetal2020} for the same data, or by \cite{kremweis2013}, for another dataset. Assuming steady health progress, here for Germany, the given intensity estimates must be interpreted as intensities averaged over cohorts and are too high for today. Differences, and hence the stroke effect, could still be adequate.

\begin{appendix}
	
	\section{\sc Lifetime State Model} \label{lsm}
	
\begin{figure}[htb!] \centering
	\setlength{\unitlength}{0.8cm}
	\begin{picture}(10,5.8)
		\linethickness{0.3mm}
		\put(0,0.3){\vector(1,0){9.0}} \put(4.0,-0.2){calender time}
		\put(0,2.2){\line(0,1){2}}  
		\put(-1.3,4.8){$\overbrace{\vphantom{ } \hspace{2.3cm}}^{\mbox{\small birth period (length 54 years)}}$}
		\put(3,2.2){\line(0,1){2}} 
		\put(7.5,2.2){\line(0,1){2}} 
		
		\linethickness{0.15mm}
		
		\put(2,3.7){\circle*{0.15}} \put(8.275,3.7){\circle{0.15}} 
		\put(1.8,3.7){ \line(1,0){5.6}}    
		\multiput(7.2,3.7)(0.2,0){5}{\line(1,0){0.1}}
		
		\put(0.5,3.3){\circle*{0.15}} \put(1.575,3.3){\circle{0.15}} 
		\multiput(0.5,3.3)(0.4,0){3}{\line(1,0){0.2}}

		\put(1.75,2.5){\circle*{0.15}} \put(5.75,2.5){\circle{0.15}} 
		\put(1.625,2.5){ \line(1,0){3.925}}

		\put(2.0,3.8){$\overbrace{\vphantom{ } \hspace{4.4cm}}^{\mbox{\small age-at-study-end $U$ + 10}}$} 
		\put(0.3,2.4){$\underbrace{\vphantom{ } \hspace{1.0cm}}_{\mbox{\small age-at-study-begin $U$}}$}
		\put(1.8,2.55){$\overbrace{\vphantom{ } \hspace{3.1cm}}^{\mbox{\small age-at-death $T$}}$}  
		
		\put(1.7,1.4){$\underbrace{\vphantom{ } \hspace{3.5cm}}_{\mbox{\small observation period (length 10 years)}}$} 
	\end{picture}
	\caption{Three cases of the date of birth (black bullet) and date of death (white circle): observed lifetime (bottom, solid), unobserved (i.e. left-truncated) person  (middle, long dashed), observed life up to study end (solid, top) and unobserved (i.e. right-censored) duration until death (short dashed, top) }
	\label{bspbild}
\end{figure}
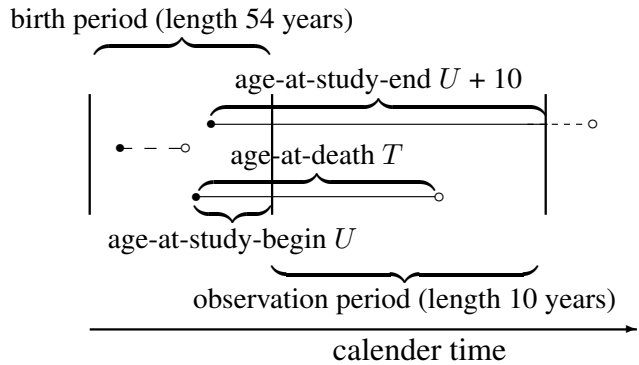

Consider a preliminary model in order to lay out the stochastic details more easily. The population is unchanged to that of the univariate model of Section \ref{dsm1} with several disease states, we aim at Germans born in the first half of the $20^{th}$ century (see first line in Figure \ref{data_selection}). Again the same HCD 2004-2013 are to be used. In the disease state model of Section \ref{dsm1}, death is only  the event of truncation, but not that of interest. To `collapsed' the `non-dead' states ($H$,$S^1$ and $D$) to one `alive' state simplifies the notation. As population model here, only of interest is the lifetime $T$ after the 50$^{th}$ birthday, called `age-at-death' in the following (see Figure \ref{bspbild}). It has hazard rate $\lambda_E(\cdot) \equiv \lambda$ (i.e. is Exponentially distributed) with CDF $F_E(\cdot)$.

For a person drawn randomly from the population, the age-at-death $T$ is defined on the probability space $(\Omega, \mathcal{F}, P_{\lambda})$. 
\begin{enumerate}[label=\AnnANumm]
	\item \label{pararaumeinfach} It is for the true parameter $\lambda_0 \in \Lambda:=[\varepsilon;1/\varepsilon]$ for some small $\varepsilon \in ]0;1[$.
\end{enumerate}

The population model is further described by a second measurement, the time elapsed for a person since the age of 50 at study begin, $U$, denoted `age at study begin' (see Figure \ref{bspbild}). It is an affine transformation of the birthdate. The distribution of $U$ will not be important.

There is no value in using a symbol for the number of years over which we observe, 2004 -- 2013 in our case study, other than 10. That number will not occur in any other meaning. 

\subsection{Filtration and conditional likelihood contribution}

It is well-known that for a simple random sample the maximum likelihood estimator for $\lambda$ is a ratio, where each person contributes a numerical one to the numerator and the time at risk, $T$, to the denominator.   

Let $t$, as in Section \ref{dsm1}, count the years after a person's 50$^{th}$ birthday. That birthday is typically the earliest age at which a stroke occurs, and we continue to simply call $t$ `age'. Of course, the age 50 is not the earliest age at which a person may die, but for simplicity we may think alike here. Methodologically, the definition of the origin of life, the birth, is arbitrary in this model here. For the model in Section \ref{dsm1}, studied the effect of left-truncation due to death before 2004, and refrain from studying the effect of truncation due to death before age 50 because death before age 50 is still rare.

The Let $X_t$ indicate a person's state at the age of $t$. For a person with dementia we write $X_t=D$, irrespective of whether a stroke has preceded dementia onset at that age or not. 
Equivalently to $T$, the experiment can be expressed in terms of a process, $N_T(t):=\mathds{1}_{\{T \le t\}}$ (for $t \in \mathbb{R}^+$). In order to prepare for modelling left-truncation, adapt it to $\mathcal{N}_t:=\sigma\{\mathds{1}_{\{T \le s\}}, 0 \le s  \le t\}$ and note that $N_T$ has a compensator with intensity $\alpha(t):= \mathds{1}_{\{N_T(t-)=0\}} \lambda=Y(t) \lambda$, with respect to $\mathcal{N}_t$ and $P_{\lambda}$ and with $Y(t):= \mathds{1}_{\{T \ge t \}}$.

\subsubsection{Non-random left-truncation and right-censoring} \label{ltrcfixed}

\begin{figure}[ht]
	\begin{center}
		\begin{picture}(400, 150)
			
			
			\put( 0.0, 15.0){\thinlines\vector(1, 0){400.0}}
			\put(0.0, 13.0){\thinlines\line(0, 1){12.0}}

			\put(333.0, 0.0){\footnotesize{01/01/2004}}
			\put(353.0, 13.0){\thinlines\line(0, 1){140.0}}
			
			\put(363.0, -10.0){\footnotesize{01/01/2014}}
			\put(383.0, 13.0){\thinlines\line(0, 1){140.0}}
			
			\put(-15.0, -4.0){\footnotesize{01/01/1950}}

			
			\put(85.0, 135.0){\thinlines\line(1, 0){285.0}}
			\put(85.0, 135.0){\thinlines\circle{2.0}}
			\put(370.0, 135.0){\thinlines\circle*{2.0}}

			\put(85.0, 137.0){$\overbrace{\textcolor[rgb]{0.93,0.93,0.93}{\hspace*{10cm}}}$}
			\put(210.0, 148.0){\normalsize{$T=34$}}

			
			\put(45.0, 90.0){\thinlines\line(1, 0){285.0}}
			\put(45.0, 90.0){\thinlines\circle{2.0}}
			\put(330.0, 90.0){\thinlines\circle*{2.0}}

			\multiput(328.0, 90.0)(3,0){9}{\thinlines\line(1, 0){1.0}}

			\put(45.0, 92.0){$\overbrace{\textcolor[rgb]{0.93,0.93,0.93}{\hspace*{10cm}}}$}
			\put(170.0, 103.0){\normalsize{$T=35$}}	
			
			\put(45.0, 87.0){$\underbrace{\textcolor[rgb]{0.93,0.93,0.93}{\hspace*{10.8cm}}}$}
			\put(170.0, 69.0){\normalsize{$u \; (\text{or} \; U)=39$}}	
			
			
			\put(50.0, 37.0){\thinlines\line(1, 0){350.0}}
			\put(50.0, 37.0){\thinlines\circle{2.0}}
			\put(400.0, 37.0){\thinlines\circle*{2.0}}

			\put(50.0, 39.0){$\overbrace{\textcolor[rgb]{0.93,0.93,0.93}{\hspace*{12.3cm}}}$}
			\put(210.0, 50.0){\normalsize{$T=50$}}	
			
			\put(50.0, 36.0){$\underbrace{\textcolor[rgb]{0.93,0.93,0.93}{\hspace*{11.7cm}}}$}
			\put(185.0, 18.0){\small{$u \;(\text{or} \; U)+10=49$}}

		\end{picture}
	\end{center}
	\caption{	(top: uncensored/untruncated) Path for person born 1/1/1925 with death 01/01/2009, i.e. with $T$ = 84 - 50, $u$  (or $U$) $\le T \le u$ (or $U$) +10 
		\newline (middle: left-truncated) Path for person born 1/1/1915 with death 01/01/2000, i.e. with $T$ = 85 - 50, $T < u$ (or $U$), $u$ (or $U$) = 89 - 50 \newline 
		(bottom: right-censored) Path for person born 1/1/1915 with death 1/1/2015, i.e. with $T$ = 100 - 50, $T \ge u$ (or $U$) + 10, $u$ (or $U$) + 10 = 99 - 50 \newline (Explanation of graphs and symbols is distributed over larger parts of text.)}
	\label{c_figure}
\end{figure}
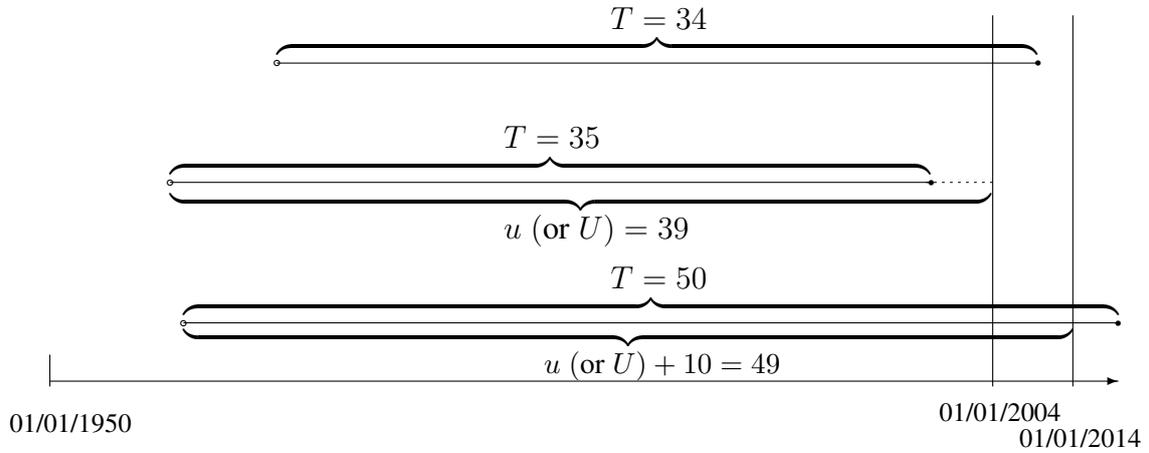

Informally, left-truncation means that, looking at the top path in Figure \ref{c_figure}, the person is only observable at risk of death from 01/01/2004 on, i.e. $T-u$ = 34 - 29 = 5 years. For simplicity, we assume $u$ to be deterministic here, i.e. the experiment to been planned. This is unrealistic for our case study and will be relaxed in Section \ref{mortalran}. Intuitively the person should now contribute to the estimator of $\lambda$, still in the numerator with a numerical one, and in the denominator no longer with $T$ but only with $T-u$. (We will see that this does not strictly maximize the likelihood.) We decide to start observation at the age of $u$ and use the probability measure of $N_T$ thereafter, namely use a marginal likelihood. The reduced observation in the stochastic process model corresponds to a coarser filtration. (Increasing the filtration will be necessary for random $U$.) 
The observable filtration is now
\begin{equation*}
	\leftindex_u{\mathcal{G}}_t  :=  \sigma\{ \mathds{1}_{\{u \le T \le s\}}, u \le s \le t\}
\end{equation*}	
and the lost observation, as compared to $\mathcal{N}_t$, can be seen from completing the filtration with $\leftindex_u{\mathcal{G}}_t:=\{ \emptyset, 0 \le s < u\}$. (If the case $T\ge u$ is considered as `retrospectively ascertained', \cite{weiswied2021} include the entire path of $N_T$ into a maximum likelihood analysis, but in a manner that is less easily extended to right-censoring.) Extending $\leftindex_u{\mathcal{G}}_t$ to censoring will end this section, and is needed in our case study.

We will now see how to proceed with the middle person in Figure \ref{c_figure}. It is interesting to note that, with whatever $u$, observing a person is indication of a small $\lambda$ and not observing a person is indication of a larger $\lambda$. Hence any left-truncated person must enter the likelihood, although we do not see it. This includes that we do not know how many there are.  Formally, if $N_T(u)=1$, no stochastic development will occur after $u$, which is formalised in $\leftindex_u{N}_T(t):=N_T(t) - N_T(\min(t,u))$, i.e. $\leftindex_u{N}_T$ is $N_T$ if $u \le T$ and constantly zero otherwise. Note that $\leftindex_u{\mathcal{G}}_t  = \sigma\{\leftindex_u{N}_T(s), u \le s \le t\}$.

Aiming at likelihood-based estimation, a density starting from $u$, i.e a marginal likelihood \cite[see][Definition 7.2(i)]{gourieroux1995} is needed. A likelihood is, with respect to some dominating measure, the Radon-Nikodym derivative of the measure that describes the experiment. (The derivative is then evaluated at the observed data.) The measure of an experiment will usually contain the location of the experiment. And for a Bernoulli-experiment, only the location is needed. Considering $\leftindex_u{N}_T$ as series of Bernoulli increments over infinitely short intervals, expectations of the increments define the intensity process. After a deterministic $u$, intuitively, the expected increase of $\leftindex_u{N}_T$ over an interval of length $dt$, at age $t$, is $\lambda dt$ if death is not reached, i.e. $t \le T$, and if the person is no left-truncated, i.e. in case of $A:=\{T > u\}$. Else it is zero. We assume $P_{\lambda}(A) > 0$. Formally, the intensity process of $\leftindex_u{N}_T(t)$ after $u$ is
$\leftindex_u{\alpha}(t):=\mathds{1}_{\{u < t \le T\}} \lambda$, 
with respect to the dominating probability measure 
\[
P^{A}_{\lambda}(F):= \frac{P_{\lambda}(F \cap A)}{P_{\lambda}(A)}, \quad \text{for} \quad F \in  \mathcal{F},
\]
that conditions on $A$. For a person not truncated (see Figure \ref{c_figure}, top) we lose information, but due to the zero-intensity of a truncated person (see Figure \ref{c_figure}, middle) the contribution to the criterion function is ineffective and hence not observing the person, still renders the criterion function observable. More formally, that the dominating measure depends on $\lambda$ can be interpreted as information loss, namely `remaining' in the dominating measure. For a proof that $\leftindex_u{N}_T$, compensated by $\int_u^t\leftindex_u{\alpha}(s) ds$, really is a $(P_{\lambda}(A),\leftindex_u{\mathcal{G}}_t)$-martingale, with deterministic $u$ as special case of the random $U$, see Proposition 4.1 in \cite{And0}. We can denote a criterion function, build on the latter intensity, a marginal likelihood. But conditioning will follow.

Right-censored is the age-at-death if it occurs after 2013, or after having left the AOK (see Figures \ref{data_selection} and \ref{c_figure}, bottom).	As in \cite[][Examples III.3.6, IV.1.7 and VI.1.4]{And} we superimpose  right-censoring on  left-truncation. The observed left-truncated and right-censored counting process is
\[
\leftindex_u{N}^c_T(t) := \int_0^t C(s) d \leftindex_u{N}_T(s)= \mathds{1}_{\{u \le T \le \min(t, u+10)\}}
\]
with $C(t):= \mathds{1}_{\{t \le u+10\}}$  \cite[compare][Example 1.4.2]{flem1991}. It has intensity
$\leftindex_u{\alpha}_T^c(t) := \lambda \mathds{1}_{\{u \le t \le \min(T,u+10) \}}$	
with respect to $P^A_{\lambda}$ and observed filtration 
\[
\leftindex_u{\mathcal{F}}^c_t := \sigma\{\leftindex_u{N}^c_T(s), u \le s \le t\}= \sigma\{\mathds{1}_{\{u \le T \le \min(s, u+10)\}},  u \le s \le t\}.
\]

The $\leftindex_u{\mathcal{F}}^c_t$ is self-exciting \cite[see][p4]{And0}, as required \cite[see][p23]{And0}, so that the density is determined by Jacod's formula \cite[see][Formula (4.3)]{And0}, \cite[see also][Formula (2.7.2'') (and in extension 3.2.8)]{And}:
\begin{eqnarray}
	dP & = & \Prodi_{u < t \le u+10} \{ (1-\leftindex_u{\alpha}_T^c(t)dt)^{1- d\leftindex_u{N}_T^c(t) } [\leftindex_u{\alpha}_T^c(t)]^{d\leftindex_u{N}_T^c(t)} \} \nonumber \\
	& = & 
	\begin{cases}
		1^1 \cdot 0^0=1  & \text{for}\; T < u \\ 
		\lambda(T) e^{- \int_u^T \lambda(t) dt}=	\lambda e^{- \lambda (T-u)}  & 
		\text{for} \; u \le T \le u + 10  \\ 
		e^{- \int_u^{u+10} \lambda(s) ds}=e^{- 10 \lambda}  & \text{for} \; u+10 < T \\
	\end{cases}   \label{lihoodexp} 
\end{eqnarray}
For an explanation of the second and third line, see page 24 and, respectively, Example 2.2 in \cite{And0}. Note that in the first line $d\leftindex_u{N}_T^c(t) \equiv 0$ and $\mathds{1}_{\{u \le T \le \min(t,u+10)\}} \equiv 0$. For whatever process $\mathbf{Z}$, one defines  $d Z(t):= Z(t)-Z(t - dt)$ for some `small' $dt$ \cite[see][Section 1.4]{flem1991}. Note that the exponential function arises similar to $e^x=\lim_{dt \to 0} (1+ x dt)^{dt^{-1}}$.

Independent persons are, due to different $u_i$, not identically distributed. Each density is a Radon-Nikodym derivative with respect to a different measure, all of which are even dependent on the parameter, $P_{\lambda}^{A_i}$.
Even worse, measures are conditional on observation. That the density for a collection of independent persons is the product of the persons' densities (see e.g. \citet[][Sect. 8.2]{Bley2021} or \citet[][Sect. III.1.Example(a)]{Fel2}), relies on the equal (and parameter-independent) dominating measure (usually being Lebesgues). To achieve equal dominating measures for all persons, we will and can follow Examples IV.1.7 and VI.1.4 of \cite{And} in using a random $U$. The sample in our case study of size {\em n}$_{all}$ implies that indeed $T$ and $U$ are random.  

\subsubsection{Random left-truncation and conditional right-censoring} \label{mortalran}

The probability space for $(T,U)$ is $(\Omega, \mathcal{F}, \tilde{P}_{\lambda})$, where the distribution of $U$ will not be important and we suppress its parameter (and indicate the difference to $P_{\lambda}$ with the tilde instead). 
\begin{enumerate}[label=\AnnANumm]
	\setcounter{enumi}{1}	
	\item \label{moeglich} $U$ and $T$ are independent and $\beta_{\lambda_0}:=\tilde{P}_{\lambda_0}(A) > 0$ with redefined $A:=\{T > U\}$. 
\end{enumerate}

\subparagraph{Left-truncation for age-at-death}

Ignoring censoring for the moment, $T$ is recorded when larger than $U$, the age-at-study-begin (see again Figure \ref{c_figure}). Consider the unobservable filtration $\mathcal{G}_t := \sigma\{ \mathds{1}_{\{T \le s\}}, \mathds{1}_{\{U \le s\}}, 0 \le s \le t\}$.
For left-truncation \cite[see][Example III.3.2]{And}, similar to non-random left-truncation, define  $\leftindex_U{N}_T(t):=N_T(t) - N_T(t \land U)$ 
with $\mathcal{G}_t$-intensity, with respect to $\tilde{P}_{\lambda}$, again being $\lambda \mathds{1}_{\{T \ge t\}}$ (because $U$ is independent of $T$). We observe durations $U$ and $T$ in the case of $A$, and neither measurement $U$ nor $T$ - nor the person at all -  when $T< U$.
Now define $\leftindex_U{Y}(t):=Y(t) \mathds{1}_{\{t > U\}}= \mathds{1}_{\{T \ge t > U\}}$,	 
and as $\mathcal{G}_t$ is unobservable, but due to $U$ being a $\mathcal{G}_t$-stopping time, $\leftindex_U{\mathcal{F}}_t:= \{ \leftindex_U{N}_T(s), \leftindex_U{Y}(s), U \le s \le t\}$ is observable. The intensity process of $\leftindex_U{N}_T(t)$ is 
$\leftindex_U{\alpha}_T(t)=\mathds{1}_{\{U < t \le T\}} \lambda$, 
with respect to $\leftindex_U{\mathcal{F}}_t$ and the probability measure 
\[
\tilde{P}^{A}_{\lambda}(F):= \frac{\tilde{P}_{\lambda}(F \cap A )}{\tilde{P}_{\lambda}(A)} \quad \text{for} \quad F \in  \mathcal{F}.
\]
See again \citet[Proposition 4.1]{And0}. Intuitively, $\leftindex_U{N} _T$ will not yet jump prior to $U$, and no longer after $T$, and in between, at the hazard rate of $T$. Now, as $\leftindex_U{\mathcal{F}}_t$ is self-exciting, we may apply Jacod's formula \cite[see][Formula (2.1)]{And0} in order to determine a conditional version of the marginal likelihood \cite[see][Formula (4.3)]{And0}. And as $U$ is independent of $T$ and the hazard rate has the form $\lambda_E(\cdot) \equiv \lambda$, the conditional (marginal) likelihood \cite[see][Definition 7.2(ii)]{gourieroux1995} is for $T \le U$ one and else \cite[see also][Formula (3.3.3)]{And} (and \citet[][p24]{And0} with $\Delta\leftindex_U{N}_T$ replaced by $d \leftindex_U{N}_T$:
\begin{eqnarray}
	\leftindex_U{L}(\lambda)	 & = & \Prodi_{t > U} \{\leftindex_U{\alpha}_T(t)^{\Delta\leftindex_U{N}_T(t)}  (1-\leftindex_U{\alpha}_T(t) dt)^{1- \Delta \leftindex_U{N}_T(t) }  \} \nonumber \\
	& = & 	\lambda_E(T) e^{- \int_U^T \lambda(t) dt} = \frac{[1- F(T)] \lambda(T)}{1- F(U)}
	= \lambda e^{- \lambda (T-U)}. 
	\label{trunclihoodexp} 
\end{eqnarray}

\subparagraph{Acknowledging non-random censoring}

First note that now $T > U+10$ is known, conditionally on $A$, because $U$ is observable, rendering unnecessary a further increase in filtration. Hence, similar to Section \ref{ltrcfixed}, we apply Definition III.2.1 in \cite{And}, with $\leftindex_U{N}_T$ in the role of $N$, $(\leftindex_U{\mathcal{F}}_t)$ in the role of $(\mathcal{F}_t)$, $\leftindex_U{\alpha}_T$ in the role of $\lambda^{\theta}$ and $\tilde{P}^{A}_{\lambda}$ in the role of $P_{\theta \psi}$. Being conditionally deterministic, $C(t):= \mathds{1}_{\{t \le U+ 10\}}$ is independent and predictable. We
define $\leftindex_U{Y}^c(t):=C(t) \leftindex_U{Y}(t)= \mathds{1}_{\{U \le t < \min(T,U+10)\}}$,
so that 
\[
\leftindex_U{N}_T^c(t):= \int_U^t C(s) \leftindex_U{N}_T(s) = \mathds{1}_{\{U \le T \le \min(t,U+10)\}}
\]
has intensity $\leftindex_U{\alpha}_T^c(t)=\mathds{1}_{\{U < t \le \min(T,U+10)\}} \lambda$, 
with respect to the observable filtration
\[
\leftindex_U{\mathcal{F}}^c_t:= \{ \leftindex_U{N}^c_T(s), \leftindex_U{Y}^c(s+), U \le s \le t\}
\]
and conditional distribution $\tilde{P}^A_{\lambda}$  \cite[see][Section III.2.2]{And}.
By Formula (3.2.8) in  \cite{And}, the partial (and here conditional marginal) likelihood is for $T \le U$ again one and otherwise:
\begin{eqnarray}
	\leftindex_U{L}_{\tau}^c(\lambda)	 & = & \Prodi_{t > U}^{\tau} \{[\leftindex_U{\alpha}^c_T(t)]^{\Delta\leftindex_U{N}^c_T(t)}  (1-\leftindex_U{\alpha}^c_T(t) dt)^{1- \Delta \leftindex_U{N}^c_T(t) }  \} \nonumber \\
	& = & 
	\begin{cases}
		\lambda(T) e^{- \int_U^T \lambda(t) dt} = \lambda e^{- \lambda (T-U)}  &  \text{for} \; U \le T \le U + 10 < \tau  \\ 
		e^{- \int_U^{U+10} \lambda(s) ds} =e^{- 10 \lambda}  & \text{for} \; U+10 < T < \tau \\
	\end{cases} \label{trunccenslihoodexp} 
\end{eqnarray}
For the second line: For whatever process $\Delta Z(t):= Z(t) - Z( t-)$ \cite[see][Sect. II.2]{And}, hence $\Delta \leftindex_U{N}^c_T(t)= \leftindex_u{N}^c_T(t) - \leftindex_u{N}^c_T(t-)$ is only not zero if $\leftindex_u{N}^c_T(t)$ jumps. These jumps are of height one, since the transition times are continuous.


\subsubsection{Large sample properties and standard error} \label{appsec4}

We now combine the contribution \eqref{trunccenslihoodexp} for each person $i$ (truncated or not) to the conditional likelihood for the simple sample of unobserved {\em n}$_{all}$. We denote the size of the observed sample, combining all persons surviving $31/12/2003$, as {\em n} := $\sum_{i=1}^{n_{all}} \mathds{1}_{A_i}$ where $i$ enumerates the persons (see Figure \ref{data_selection}). Without loss of generality, we have sorted those not truncated at the beginning of the sample, and all other factors are one (see again the first line of \eqref{lihoodexp}). 
Due to the assumption of a simple sample (and hence the same dominating measure), the logarithmic conditional likelihood for the data (we suppress marginalisation from now on) is the sum of logarithms for \eqref{trunccenslihoodexp}. We denote by  {\em N}$_{cens}$ := $\sum_{i=1}^n \mathds{1}_{\{T_i > U_i+10\}}$ the number of right-censored and by {\em N}$_{uncens}$ := {\em n} - {\em N}$_{cens}$ = $\sum_{i=1}^n \mathds{1}_{\{U_i < T_i \le U_i+10\}}$ the number of neither truncated nor censored people and can write $\ln \leftindex_U{\mathbf{L}}_{\tau}^c(\lambda)$ as 
\begin{equation} \label{loglikemodela}
	\sum_{i=1}^n \ln \leftindex_U{L}_{\tau}^c(\lambda)_i  = N_{uncens} \ln(\lambda)  - \lambda \sum_{i=1}^{n} \mathds{1}_{\{U_i \le T_i \le U_i +10 \}} (T_i- U_i) -  10 \lambda N_{cens}, 
\end{equation}
so that the unique estimator (if in $\Lambda$, see \ref{pararaumeinfach}) becomes
\begin{equation}\label{expest}
	\hat{\lambda} =argmax_{\lambda \in \Lambda} \ln \leftindex_U{\mathbf{L}}_{\tau}^c(\lambda) =    \frac{N_{uncens}}{\sum_{i=1}^{n} \mathds{1}_{\{U_i \le T_i < U_i+10\}} (T_i- U_i) + 10 N_{cens}}.	
\end{equation}
For our parametric model, we assume: 
\begin{enumerate}[label=\AnnANumm]
	\setcounter{enumi}{2}
	\item \label{varpos2} $\int_0^{\tau}  \tilde{P}_{\lambda}(U < t \le \min(T,U+10)| A)dt > 0$.
\end{enumerate}

\begin{theorem}\label{normmodela} Under Conditions \ref{pararaumeinfach}-\ref{varpos2} it is $\hat{\lambda}$ of \eqref{expest} consistent and $\sqrt{n_{all}}(\hat{\lambda} - \lambda_0) \stackrel{\mathcal{D}}{\longrightarrow} \mathcal{N}(0,\sigma^{-1}(\lambda_0))$, for $n_{all} \to \infty$, with
	\[
	\sigma(\lambda_0): = \frac{\beta_{\lambda_0}}{\lambda_0} \int_0^{\tau} \tilde{P}_{\lambda_0}(U < t \le \min(T,U+10)| T > U) dt.
	\]
\end{theorem}

{\bf Proof}:
Due to the uniqueness of $\hat{\lambda}$, for both consistency and asymptotic normality, we need to verify Conditions (A)-(E) \cite[see][Theorems VI.1.1+2]{And}. In order to map the notations, note that {\em n} becomes {\em n}$_{all}$, $a_n:= \sqrt{n_{all}}$,  $\boldsymbol {\theta}$ becomes $\lambda$ and  $h$ is redundant. The $\lambda(t;\theta)$ becomes, by interchanging conditional expectations with summation, the multiplicative intensity process of $\leftindex_U{N}^c_{T\bullet}(t):= \sum_{i=1}^n \leftindex_U{N}^c_{Ti}(t)$:
\begin{equation*}\label{multiintensity11111} 
	\leftindex_U{\alpha}^c_{\bullet}(t;\lambda)= \lambda \sum_{i=1}^n \mathds{1}_{\{U_i < t \le \min(T_i,U_i+10)\}}
\end{equation*}

By \ref{pararaumeinfach}, (A) is fulfilled for intensity $\leftindex_U{\alpha}_T^c(t)$ and therefore for $\leftindex_U{\alpha}_{T\bullet}^c(t;\lambda)$ and the logarithm of likelihood \eqref{trunccenslihoodexp}. For (B),  
because by the LLN, for fixed $t \in [0,\tau]$, for the portion in the study period alive at the age $t$ and observed it is
\begin{equation} \label{111}
	\frac{1}{n}\sum_{i=1}^n \mathds{1}_{\{U_i < t \le \min(T_i,U_i+10)\}} \stackrel{P}{\longrightarrow} \tilde{P}_{\lambda} (U < t \le \min(T,U+10)| A),
\end{equation} 
due to the simple sample assumption.
Furthermore $n/n_{all} 	\stackrel{P}{\longrightarrow} 	\tilde{P}_{\lambda_0}(A)=\beta_{\lambda_0}$, so that Slutzky's Lemma and the CMT yield
\begin{multline*}
	\frac{1}{n_{all}} \int_0^{\tau} \left(\frac{d}{d \lambda} (\ln \lambda) \right)^2|_{\lambda=\lambda_0} \lambda_0 \sum_{i=1}^n \mathds{1}_{\{U_i < t \le \min(T_i,U_i+10)\}} dt \\
	=  \frac{n}{n_{all}} \frac{1}{n} \sum_{i=1}^n \int_0^{\tau} \frac{1}{\lambda_0^2}  \lambda_0  \mathds{1}_{\{U_i < t \le \min(T_i,U_i+10)\}} dt 
	\rightarrow  \sigma(\lambda_0). 
\end{multline*}
For (C), because (i) {\em n} $\le$ {\em n}$_{all}$, (ii) $1/(n \lambda_0) \stackrel{P}{\rightarrow} 0$ by \ref{pararaumeinfach} and $n \stackrel{P}{\rightarrow} \infty$ (due to {\em n} following a Binomial distribution with parameters $n_{all}$ and the selection probability $\beta_{\lambda_0}$) and (iii) \ref{varpos2}, we have
\begin{multline} \label{mitteleinfach}
	\frac{n}{n_{all}}	\frac{1}{n}	\int_0^{\tau} \left(\frac{d}{d \lambda} (\ln \lambda)|_{\lambda=\lambda_0}\right)^2 \mathds{1}_{\{|\frac{1}{n} \frac{d}{d \lambda} (\ln \lambda)|_{\lambda=\lambda_0}| > \varepsilon \}} \lambda_0 \sum_{i=1}^n \mathds{1}_{\{U_i < t \le \min(T_i,U_i+10)\}}  dt 	\\
	= \frac{n}{n_{all}}	\frac{1}{\lambda_0^2 } \mathds{1}_{\{|1/(n \lambda_0) | > \varepsilon \}} \lambda_0 	\int_0^{\tau} \frac{1}{n} \mathds{1}_{\{U_i < t \le \min(T_i,U_i+10)\}}  dt  
	\stackrel{P}{\rightarrow} 0.
\end{multline}
For (D), by \ref{varpos2}, it is $\sigma(\lambda_0)$ positive. Condition (E) consists of six conditions: For the first four, note that 
\begin{multline*}
	\sup_{\lambda} \left| \frac{\partial^3}{d \lambda^3} \left(\lambda \sum_{i=1}^n \mathds{1}_{\{U_i < t \le \min(T_i,U_i+10)\}}\right)\right| =0, \\
	sup_{\lambda} \left|\frac{\partial^3}{d \lambda^3} \ln(\lambda)\right| =  sup_{\lambda} |\lambda^{-3}/2| =: H < \infty \quad \text{and} \quad
	\frac{1}{n_{all}} \int_0^{\tau} H dt \rightarrow 0. 
\end{multline*}
For the fifth, note \eqref{mitteleinfach} (and also \ref{pararaumeinfach}). For the sixth, note  \eqref{mitteleinfach} and then $H/\sqrt{n_{all}} 	\stackrel{P}{\rightarrow} 0$.

\hfill \qed

It remains to consistently estimate $\sigma(\lambda_0)$, in order to construct a confidence interval: 
\begin{eqnarray}
	- \mathcal{J}_{\tau}(\lambda_0) & := & \frac{d^2}{d \lambda^2} \ln \leftindex_U{\mathbf{L}}_{\tau}^c(\lambda)|_{\lambda=\lambda_0} 
	\stackrel{\text{\eqref{loglikemodela}}}{=}  - \frac{N_{uncens}}{\lambda_0^2} \nonumber \\
	\frac{1}{n_{all}} \mathcal{J}_{\tau}(\lambda_0) & \stackrel{n_{all} \to \infty, P}{\longrightarrow} &  \sigma(\lambda_0) \quad \text{\cite[][Formula (6.1.11)]{And}} \nonumber \\	
	\Rightarrow \mathcal{J}_{\tau}^{-1}(\lambda_0) & = & \frac{\lambda_0^2}{N_{uncens}} \quad \text{and} \quad n_{all} \mathcal{J}_{\tau}^{-1}(\lambda_0) \stackrel{\cdot}{=}  \sigma^{-1}(\lambda_0) \nonumber \\
	\Rightarrow Var(\hat{\lambda}) & \stackrel{\cdot}{=} & \mathcal{J}_{\tau}^{-1}(\hat{\lambda}) \quad \text{(by \eqref{expest} and CMT as $\hat{\lambda}$ is consistent)}  \nonumber\\ & = &  \frac{N_{uncens}}{\left(\sum_{i=1}^{n} \mathds{1}_{\{U_i \le T_i < U_i+10\}} (T_i- U_i) + 10 N_{cens}\right)^2} \label{expstand}
\end{eqnarray} 
Note that even though 	$Var(\hat{\lambda}) = Var (\sqrt{n_{all}}\hat{\lambda})/n_{all}\stackrel{\cdot}{=}\sigma^{-1}(\lambda_0)/n_{all}$, by Theorem \ref{normmodela}, $Var(\hat{\lambda})$ formally depends on the unobservable  {\em n}$_{all}$, the standard error of $\hat{\lambda}$ (SE), as the square root of \eqref{expstand}, is still observable.  


\subsubsection{Finite sample properites} \label{simmort}

Comparable to Section \ref{finpropdsm}, we visualise consistency and asymptotic normality, stated in Theorem \ref{normmodela}. We again find the approximations suitable for the AOK HCD data of {\em n} $\approx$ 250{,}000 observations in the next Section \ref{case1}. 

For an appropriate parameter, similar to Section \ref{sec232}, for the Exponential distribution of $T$, note first that $P(T \le 1)$ = 1 - $e^{-\lambda_0}$ $\approx$  $\lambda_0$ for small $\lambda_0$ (by the well-known $e^x$ $\approx$ 1 + {\em x}, for small {\em x}). For a contemporaneous one-year death incidence (and hence intensity) in Germany, $\approx$ 0.96 mio. deaths in 2020, restricted to the over 50 year-old's  \cite{sterbe}, are to be set in relation to $\approx$ 35.5  mio. alive over 50 year old's in 2019 \cite{bestand}, i.e. $F_E(1) \approx \lambda_0$ $\approx$ 0.035. The point estimate  in the case study (as of \eqref{expest}) will be $\hat{\lambda}$ = 0.022 and because smaller parameters are more difficult to estimate we use $\lambda_0$ = 0.02. 

In order to mimic the portion of uncensored in our case study, we simulate, independent of $T$, $U \sim Exp(0.004)$, so that (on average) from {\em n} = 250,000 observations {\em n}$_{uncens}$ = 40,000 are uncensored.
For different sample sizes and 2,000 simulated studies each, Table \ref{msemort} shows that the mean square error decreases with increasing sample size as to be expected and reaches an irrelevant magnitude far below the sample size of our case study. 
\begin{table}[h]
	\centering
	\caption{Mean squared error (MSE) for estimator \eqref{expest} of the death hazard ($\times$ 10$^2$)}
	\label{msemort} \smallskip
	{\footnotesize
		\begin{tabular}{cccc}
			\hline
			\noalign{\smallskip}
			{\em n}$_{all}$ = 100 &{\em n}$_{all}$ = 1,000 & {\em n}$_{all}$ = 10,000 & {\em n}$_{all}$ = 100,000 \\ 	\hline	
			\noalign{\smallskip}
			0.017 & 0.0016  & 0.00016 & 0.000016 \\  \hline
			\noalign{\smallskip}       
	\end{tabular}}
\end{table}
For the assessment of normality, the {\em n} = 250,000 observations in our case study correspond to {\em n}$_{all}$ = 1.5 million sampled units. The plot of the kernel smoothed histogram is in Figure \ref{simmorb} (right panel) and confirms the normal shape.


\subsection{Result for AOK HCD} \label{case1}

The population, the sample of size {\em n}$_{all}$, and among those, the {\em n} = 236,039 people not truncated, are described in Section  \ref{res2}. The conditional likelihood contributions are \eqref{loglikemodela}. 
\begin{table}[b]
	\centering
	\caption{Statistics (as counts $\#$, or in years) for point estimate \eqref{expest} and standard error (SE) by \eqref{expstand}  for death hazard, $\sum_{i=1}^{n_{uncens}}$ requires corresponding ordering}
	\label{statmort} \smallskip
	{\footnotesize
		\begin{tabular}{ccccc}
			\hline
			\noalign{\smallskip}
			$\#$ `Survivors' & $\#$ `Deaths' & `Time at Risk of observed Deaths' & Point & SE \\ 
			{\em n}$_{cens}$ & {\em n}$_{uncens}$ & $\sum_{i=1}^{n_{uncens}} \mathds{1}_{\{u_i \le t_i < u_i+10\}} (t_i- u_i)$  & $\hat{\lambda}$ & $\sqrt{\mathcal{J}_{\tau}^{-1}(\hat{\lambda})}$ \\	\hline	
			\noalign{\smallskip}
			171,617 & 64,442  & 110,940 & 0.0353 & 0.00014 \\  \hline
			\noalign{\smallskip}       
	\end{tabular}}
\end{table}

For the ease of argumentation, we hold up the assumption that a person cannot die before age 50. Within the observation period 2004 until 2013, {\em n}$_{uncens}$ = 64,442 persons, out of the $n$, die.  We monitor histories for transitions up to 31/12/2013, so that the maximally observed timespan is $\tau$ := 54 + 10 = 64 years (after a person's 50$^{th}$ birthday). For the Exponentially distributed age-at-death, $T$, according to \eqref{expest}, point estimation of $\lambda_0$ divides {\em n}$_{uncens}$ by the observed time as risk. Each person surviving 2013 has been 10 years at risk in the observation period. For an observed death, the person contributes with the residual lifetime to the time at risk. The two contributions are listed in Table \ref{statmort} and add up to 1.8 million years, i.e. $\hat{\lambda}$ = 0.035. Note that the person-years at dementia risk in Table \ref{desstat} are by definition smaller. Recall that neither for the point estimate \eqref{expest} nor for its standard error \eqref{expstand} is the knowledge of {\em n}$_{all}$ necessary. Theorem \ref{normmodela} yields the confidence interval $\hat{\lambda} \pm z_{1 - 0.025} \sqrt{\sigma^{-1}(\lambda_0)}/\sqrt{n_{all}}$, with 97.5\% quantile of the standard Gaussian distribution z$_{1 - 0.025}$, being very narrow due to small standard error $\mathcal{J}_{\tau}^{-1/2}(\hat{\lambda})= [n_{all} \sigma(\hat{\lambda})]^{-1/2}$ by \eqref{expstand} (see again Table \ref{statmort}). The expected lifetime is 1/0.035 $\approx$ 28 (plus 50) years. That may not be appropriate, due to the assumption of a constant hazard and by the impossibility of death before age 50 in this model. Age-inhomogeneous increase in the hazard is the well-documented and in basic demography usually modelled as a Gompertz distribution. Piecewise constant hazards would be as in Section \ref{dsm2}.

	\section{\sc Proof of Theorem \ref{normmodelb}} \label{appb}
	
Due to the uniqueness of $\hat{\boldsymbol{\lambda}}$, for both consistency \cite[see][Theorem VI.1.1]{And} and asymptotic normality \cite[see][Theorem VI.1.2]{And}, we need to verify Conditions (A)-(E) \cite[see][Condition VI.1.1]{And}. Specifically, in order to map the notations, note that {\em n} becomes {\em n}$_{all}$, $a_n:= \sqrt{n_{all}}$, $\boldsymbol{\theta}$ becomes $\boldsymbol{\lambda}$, {\em h} becomes {\em hj} and $\lambda_h(t;\boldsymbol{\theta})$ becomes 
\begin{equation}\label{multiintensity} 
	\leftindex_U{\alpha}^c_{hj\bullet}(t;\boldsymbol{\lambda})=\lambda_{hj} \leftindex_U{Y}^c_{h\bullet}(t),\ hj \in \mathcal{I},
\end{equation}
bacause by interchanging conditional expectations, the multiplicative intensity process of $\leftindex_U{N}^c_{hj\bullet}(t)$
is the sum of the compensators for each person's counting process.

The fulfilment of (A) is now as in Theorem \ref{normmodela} of Appendix \ref{appsec4}. 
For (B), note first that for mixed derivatives, terms are only non-zero when $h_1j_1=h_2j_2=h_3j_3$ due to \eqref{multiintensity}. In that case,  
\begin{multline*}
	\frac{1}{n_{all}} \int_0^{\tau} \sum_{hj \in \mathcal{I}} \left(\frac{\partial}{\partial \lambda_{hj}} \ln \leftindex_U{\alpha}_{hj\bullet}^c(t;\boldsymbol{\lambda}_0) \right)^2  \leftindex_U{\alpha}_{hj\bullet}^c(t;\boldsymbol{\lambda}_0) dt \\
	=  \frac{n}{n_{all}} \frac{1}{n} \int_0^{\tau} \sum_{hj} \frac{1}{\lambda_{hj^0}^2} \lambda_{hj^0} \leftindex_U{Y}^c_{h\bullet}(t) dt 
	\rightarrow \sigma_{hj,hj}(\boldsymbol{\lambda}_0), 
\end{multline*}
because, by the LLN, for fixed $t \in [0,\tau]$, (including Slutzky's Lemma, the CMT, and \eqref{mittel} (as in Theorem \ref{normmodela})) it is $n/n_{all} \stackrel{P}{\longrightarrow}	\beta_{\boldsymbol{\lambda}_0} >0$ by \ref{posobsprobmodelb}. 
For (C), 
\begin{multline*}
	\frac{1}{n_{all}} \int_0^{\tau} \sum_{hj \in \mathcal{I}} \left(\frac{\partial}{\partial \lambda_{hj}} \ln \leftindex_U{\alpha}_{hj\bullet}^c(t;\boldsymbol{\lambda}_0) \right)^2 \\ \mathds{1}_{\left\{\left|\frac{1}{\sqrt{n_{all}}} \frac{\partial}{\partial \lambda_{hj}} \ln \leftindex_U{\alpha}_{hj\bullet}^c(t;\boldsymbol{\lambda}_0)\right| > \varepsilon\right\}} \leftindex_U{\alpha}_{hj\bullet}^c(t;\boldsymbol{\lambda}_0) dt \\
	=  \frac{1}{n_{all}} \int_0^{\tau} \sum_{hj \in \mathcal{I}} \frac{1}{\lambda_{hj^0}^2} \mathds{1}_{\left\{\left|\frac{1}{\sqrt{n_{all}}} \frac{1}{\lambda_{hj^0}}\right| > \varepsilon\right\}}\lambda_{hj^0} \leftindex_U{Y}^c_{h\bullet}(t) dt 
	\rightarrow 0,
\end{multline*}
because $1/\sqrt{n_{all}} \to 0$ and $\lambda_{hj^0}>0$ due to \ref{parraummodb}. 
For (D), $\boldsymbol{\Sigma}(\boldsymbol{\lambda}_0)$ is diagonal with $\sigma_{hj,hj}(\boldsymbol{\lambda}_0)>0$ due to \ref{posobsprobmodelb} and \ref{varpos}.
Now in (E), generally (first line) and if $h_1j_1=h_2j_2=h_3j_3$ (second line), it is:  
\begin{eqnarray*}
	\sup_{\boldsymbol{\lambda} \in \boldsymbol{\Lambda}} \left| \frac{\partial^3}{\partial \lambda_{h_1j_1} \partial \lambda_{h_2j_2} \partial \lambda_{h_3j_3} } \lambda_{hj} \leftindex_U{Y}^c_{h\bullet}(t) \right| & = & 0  \\
	\sup_{\boldsymbol{\lambda} \in \boldsymbol{\Lambda}} \left| \frac{\partial^3}{\partial \lambda_{h_1j_1} \partial \lambda_{h_2j_2} \partial \lambda_{h_3j_3} } \ln(\leftindex_U{\alpha}_{hj\bullet}^c(t;\boldsymbol{\lambda}))\right| & = & \sup_{\lambda_{hj} \in \Lambda_{hj}} \frac{1}{2 \lambda_{hj}^3}
\end{eqnarray*}
Now $1/(2 \lambda_{hj}^3) \le 1/(2\varepsilon_{hj^0}^3)< \infty$ by \ref{parraummodb}. (This explains the third and forth requirement.) For the fifth, 
\begin{multline*}
	\frac{1}{n_{all}} \int_0^{\tau} \sum_{hj \in \mathcal{I}} \left(\frac{\partial^2}{\partial \lambda_{hj}^2} \ln \leftindex_U{\alpha}_{hj\bullet}^c(t;\boldsymbol{\lambda}_0) \right)^2  \leftindex_U{\alpha}_{hj\bullet}^c(t;\boldsymbol{\lambda}_0) dt \\
	=  \frac{n}{n_{all}} \frac{1}{n} \int_0^{\tau} \sum_{(h,j)} \frac{1}{\lambda_{hj^0}^4} \lambda_{hj^0} \leftindex_U{Y}^c_{h\bullet}(t) dt 
	\rightarrow \frac{\sigma_{hj,hj}(\boldsymbol{\lambda}_0)}{\lambda_{hj^0}^2}. 
\end{multline*}
The latter are finite due to \ref{parraummodb} and \ref{varpos}. The sixth is fulfilled with the same argument as in (C).
\hfill \qed

\end{appendix}

\textbf{Acknowledgment}:
The financial support from the Deutsche Forschungsgemeinschaft (DFG)
of R. Wei\ss bach and G. Doblhammer is gratefully acknowledged (Grant 386913674 `Multi-state, multi-time, multi-level analysis of health-related demographic events: Statistical aspects and applications'). For discussions at an earlier stage of the study, we are grateful to O. Gefeller and two anonymous referees. For support with the data we thank the AOK Research Institute (WIdO) and for literature research we thank E. Rakusa. The linguistic and idiomatic advice of B. Bloch is also gratefully acknowledged. No author has any financial or commercial conflict of interest. The figure and tables haven been computed using R and Stata.

\textbf{Declarations}:
The authors declare that they have no conflict of interest.

\end{document}